\batchmode
\documentclass[english,british]{jfm}
\usepackage[T1]{fontenc}
\usepackage[latin9]{inputenc}
\usepackage{color}
\usepackage{babel}
\usepackage{array}
\usepackage{booktabs}
\usepackage{textcomp}
\usepackage{bm}
\usepackage{amstext}
\usepackage{amssymb}
\usepackage{wasysym}
\usepackage{graphicx}
\usepackage{esint}
\usepackage[unicode=true,
 bookmarks=false,
 breaklinks=false,pdfborder={0 0 1},backref=false,colorlinks=true]
 {hyperref}
\hypersetup{
 citecolor=blue}

\makeatletter

\providecommand{\tabularnewline}{\\}

\usepackage{graphicx}
\usepackage{natbib}
\ifCUPmtlplainloaded \else
  \checkfont{eurm10}
  \iffontfound
    \IfFileExists{upmath.sty}
      {\typeout{^^JFound AMS Euler Roman fonts on the system,
                   using the 'upmath' package.^^J}%
       \usepackage{upmath}}
      {\typeout{^^JFound AMS Euler Roman fonts on the system, but you
                   dont seem to have the}%
       \typeout{'upmath' package installed. JFM.cls can take advantage
                 of these fonts,^^Jif you use 'upmath' package.^^J}%
       \providecommand\upi{\pi}%
      }
  \else
    \providecommand\upi{\pi}%
  \fi
\fi

\ifCUPmtlplainloaded \else
  \checkfont{msam10}
  \iffontfound
    \IfFileExists{amssymb.sty}
      {\typeout{^^JFound AMS Symbol fonts on the system, using the
                'amssymb' package.^^J}%
       \usepackage{amssymb}%

      }{}
  \fi
\fi

\ifCUPmtlplainloaded \else
  \IfFileExists{amsbsy.sty}
    {\typeout{^^JFound the 'amsbsy' package on the system, using it.^^J}%
     \usepackage{amsbsy}}
    {}
\fi

\newcommand\Real{\mbox{Re}} 
\newcommand\Imag{\mbox{Im}} 

\newsavebox{\astrutbox}
\sbox{\astrutbox}{\rule[-5pt]{0pt}{20pt}}

\newenvironment{lyxlist}[1]
{\begin{list}{}
{\settowidth{\labelwidth}{#1}
 \setlength{\leftmargin}{\labelwidth}
 \addtolength{\leftmargin}{\labelsep}
 }}
{\end{list}}

\newcommand{\ex}[0]{\mathrm{e}}
\newcommand{\ic}[0]{\mathrm{i}}
\newcommand{\dd}[0]{\mathrm{d}}
\newcommand{\dps}[0]{\displaystyle}

\newcommand{\vs}[0]{\vspace{2mm}}

\newcommand{\uu}[0]{\underline{$u$}}
\newcommand{\uv}[0]{\underline{$v$}}
\newcommand{\uw}[0]{\underline{$w$}}
\newcommand{\up}[0]{\underline{$p$}}

\newcommand{\uum}[0]{\underline{u}}
\newcommand{\uvm}[0]{\underline{v}}
\newcommand{\uwm}[0]{\underline{w}}
\newcommand{\upm}[0]{\underline{p}}

\newcommand{\ddx}[0]{\partial_{x}}
\newcommand{\ddy}[0]{\partial_{y}}
\newcommand{\ddz}[0]{\partial_{z}}
\newcommand{\ddt}[0]{\partial_{t}}
\newcommand{\ddxx}[0]{\partial_{xx}}
\newcommand{\ddyy}[0]{\partial_{yy}}
\newcommand{\ddzz}[0]{\partial_{zz}}

\usepackage{wasysym}

\makeatother

\begin{document}

\title[Streaming in Faraday waves]{Streaming patterns in Faraday waves}

\author[N. Périnet, P. Gutiérrez, H. Urra, N. Mujica and L. Gordillo]{Nicolas Perinet$^{1}$, Pablo Gutiérrez$^{1}$, Héctor Urra$^{2}$,
\\
Nicolás Mujica$^{1}$ and Leonardo Gordillo$^{1,3}$\thanks{lgordill@ing.uchile.cl}}

\affiliation{$^{1}$Departamento de Física, Facultad de Ciencias Físicas y Matemáticas,
Universidad de Chile, Casilla 487-3, Santiago, Chile\\
$^{2}$Instituto de Física, Pontificia Universidad Católica de Valparaíso,
Avenida Brasil, Valparaíso, Casilla 2950, Chile\\
$^{3}$Department of Chemical Engineering and Materials Science, University
of Minnesota, Minneapolis, Minnesota 55455, USA}

\date{?; revised ?; accepted ?. - To be entered by editorial office}

\pubyear{2015}

\volume{??}

\pagerange{??---??}
\maketitle
\begin{abstract}
Waves patterns in the Faraday instability have been studied for decades.
Besides the rich dynamics that can be observed on the waves at the
interface, Faraday waves hide beneath them an elusive range of flow
patterns ---or streaming patterns--- which have not been studied in
detail until now. The streaming patterns are responsible for a net
circulation in the flow which are reminiscent of convection cells.
In this article, we analyse these streaming flows by conducting experiments
in a Faraday-wave setup. To visualize the flows, tracers are used
to generate both trajectory maps and to probe the streaming velocity
field via Particle Image Velocimetry (PIV). We identify three types
of patterns and experimentally show that identical Faraday waves can
mask streaming patterns that are qualitatively very different. Next
we propose a three-dimensional model that explains streaming flows
in quasi-inviscid fluids. We show that the streaming inside the fluid
arises from a complex coupling between the bulk and the boundary layers.
This coupling can be taken into account by applying modified boundary
conditions in a three-dimensional Navier-Stokes formulation for the
streaming in the bulk. Numerical simulations based on this theoretical
framework show good qualitative and quantitative agreement with experimental
results. They also highlight the relevance of three-dimensional effects
in the streaming patterns. Our simulations reveal that the variety
of experimental patterns is deeply linked to the boundary condition
at the top interface, which may be strongly affected by the presence
of contaminants along the surface.\end{abstract}

\section{Introduction}

In 1831, Michael Faraday published an article that became fundamental
for the communities of fluid dynamics and nonlinear physics \citep{1831RSPT..121..299F}.
It presented what nowadays is called the Faraday instability: the
destabilisation of the free surface of a liquid into regular ripples
after forcing its container to periodically vibrate in the vertical
direction above some threshold. Periodical vibrations introduce a
parametrical modulation to gravity, which creates regular stationary
waves on the fluid surface that oscillate at half the forcing frequency
(subharmonic waves) ---the so called Faraday waves.

It may be surprising that such a seminal contribution was only an
appendix in the 1831 article. The main content was devoted to the
study of how solid particles move on the surface of a vibrated elastic
plate and why they tend to slowly cumulate in some specific regions.
His aim was to better understand these patterns, also known as Chladni
figures \citep{Chladni}. In this context, the first goal of the experiments
included in the appendix was to change the density of the fluid surrounding
the moving particles. Regardless the nature of the fluid, a key step
to understand particle motion is to probe the velocity induced by
the oscillatory motion of the fluid.

When considering Faraday waves, the velocity field has two clearly
distinguishable components. First, there is an oscillatory part (see
the illustrative images by \citealp{wallet1950trajectoires}, reproduced
by \citealp{van1982album}), whose importance was well known for surface
waves since \citet{Stokes:1847uv} and whose linear and non-linear
mechanisms have been deeply studied \citep[see][and references therein]{Lamb2006hydrodynamics,1990AnRFM..22..143M}.
This early understanding was possible, in part, because viscous effects
are not required to describe the oscillatory part with a high level
of accuracy. Viscosity, however, is fundamental to describe the second
component of the velocity field that becomes noticeable only after
long times \citep{LonguetHiggins:1953ue}. This time-independent velocity
component was readdressed much more recently \citep{1990JFM...221..383D},
and followed by some theoretical descriptions \citep{2001PhyD..154..313V,2002JFM...467...57M,2005JFM...546..203M}.
Mean flows are crucial to understand advection of material inside
the fluid over time scales longer than the period of container oscillation.
Although the phenomenon is well known in acoustics \citep[see][ and references therein]{2001AnRFM..33...43R},
it has received little attention in the context of subharmonic waves.
We will refer to these mean flows as Faraday streaming flows or, simply,
streaming flows and concentrate our attention on them. 

The already great impact of Faraday\textquoteright s contribution
can be even extended when considering the Faraday streaming flow.
In particular, the motion produced by an oscillating fluid surface
is a relevant open question. The dispersion of pollutants on the free
surface of a flow involves several physical processes, including streaming
flows. Various processes have been considered in recent experimental
works \citep{Falkovich:2005ei,Sanl:2014hs,GutierrezAumaitre_2016b},
but a clear evaluation of the dominant effects is still lacking. On
the other hand, Faraday waves have been proposed as a way to generate
particulate films by deposition of heavy particles \citep{Wright:2003ct}
and suspended templates of light particles \citep{Chen:2014bc}, where
again streaming flows are one key for the arising patterns. In another
context, Faraday streaming flows are claimed to play an important
role on the dynamics of localised structures, namely in their drift
and interaction \citep{2001PhyD..154..313V,2002JFM...467...57M}.
Because of the variety and relevance of its applications, a thorough
study of the streaming flow (experimental, theoretical and numerical)
is needed.

In this article, we perform a detailed analysis of the Faraday streaming
flow that includes its experimental characterisation and a general
theoretical development. By performing numerical simulations, we compare
the experimental findings with the theory. In \S\ref{sec:Experiments}
we present the experiment. We briefly describe the setup and the Particle
Image Velocimetry (PIV) measurements. Then, we focus on the observed
mean flows. Section \ref{sec:Theory} concerns the theoretical development
and includes a detailed extension to three dimensions of the Batchelor
theory of oscillatory boundary layers \citep{Batchelor2000an_introduction}.
Special care was taken to present the theory in a self-consistent
way. Our analysis ends with the boundary conditions that are used
to perform simple numerical simulations in three dimensions whose
results are presented in \S\ref{sec:Numerical-simulations}. Section
\ref{sec:Discussion-2} is devoted to discussion and conclusions.

\section{Measurements\label{sec:Experiments}}

\subsection{Experimental setup\label{sub:Experimental-setup}}

To run our experiments, we manufactured a plexiglas trough of $L_{x}=190.5\,\mathrm{mm}$
long, $L_{y}=25.4\,\mathrm{mm}$ broad and $69.6\,\mathrm{mm}$ deep.
An aluminium piece attached to the top couples the trough to an electromechanical
shaker (Dynamic Systems VTS-80). A function waveform generator (Rigol
DG-1022) and a bipolar amplifier (NF HSA-4011) generate a sinusoidal
signal with tuneable amplitude and frequency that properly feeds the
shaker. The system's acceleration is precisely measured using an accelerometer
(PCB Piezotronics 340A65) and a lock-in amplifier (Stanford Research
SR830), synchronised with the shaker input signal. A second channel
of the function waveform generator is used to create a synchronisation
signal that triggers the imaging system at a programmable phase.

\begin{figure}
\begin{centering}
\includegraphics{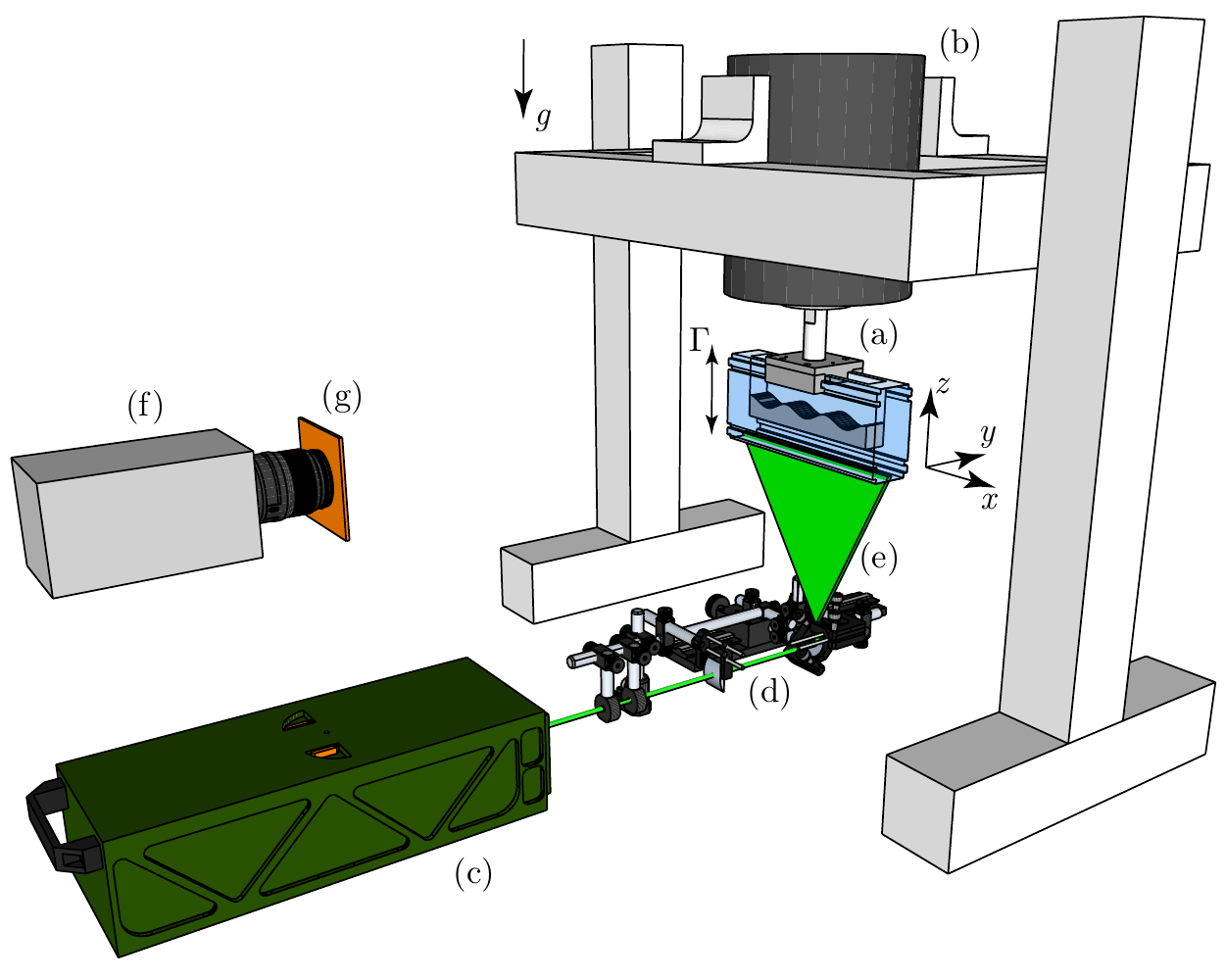}
\par\end{centering}

\protect\caption{Experimental setup. A plexiglas trough (a) is vertically driven by
an electromagnetic shaker (b). A double-pulsed laser (c) and an optical
array (d) generates a light sheet (e) which passes through the trough.
A high-speed camera (f) and a long-pass filter (g) acquires the images
for PIV. \label{fig:setup}}
\end{figure}

The upside configuration chosen for the experiment allows the system
to be illuminated from the bottom, which is the only way to avoid
undesired light refractions in Particle Image Velocimetry (PIV). A
scheme of the setup is displayed in figure~\ref{fig:setup}. Further
details can be found in \citet{Gordillo:2014jf}. The PIV imaging
system consists of a double-pulsed ND:YAG laser (Quantel Evergreen
70) and a high-speed camera (Phantom v641), properly synchronised.
The laser beam passes through an optical system that projects a $2\,\mathrm{mm}$-thick
laser sheet into the trough. The optical system, mounted on a translational
stage, allows to freely adjust the position of the illuminated plane
in the $y$ direction.

In all our runs, we filled the trough up to $h=20\,\mathrm{mm}$ with
an aqueous solution of potassium bromide (KBr). The concentration,
$13.7\%$, is such that the density of the solution matches the density
of the seeding particles: Kanomax Fluostar, $\diameter=15\,\mathrm{\text{\textmu m}},$
$\rho=1.1\,\mathrm{g/cm^{3}}$ \citep[cf. tables for aqueous solutions in][]{Lide2004Handbook}.
The seeding particles are fluorescent, a feature which is used to
create sharp images. A long-pass filter placed in front of the objective
lens filters off any reflection of the laser sheet on the free surface
and only the light emitted by the fluorescent particles reaches the
camera sensor. A small amount of wetting agent, $2\,\mathrm{ml}$
of Kodak Photoflo, is also added to the solution. The inclusion of
this agent has been shown to improve wall wettability in this kind
of system \citep{1984PhRvL..52.1421W}. Some ethyl alcohol (less than
$1\,\mathrm{ml}$) was used to previously dissolve the PIV seeding
particles into water, avoiding the formation of clusters. The mixture
of seeding particles was then kept in a syringe and added manually
to the water until the concentration of particles in images is suitable
for PIV (4 particles in an $8\times8$-pixels window, \citealt{Raffel2007Particle}).

Faraday waves are generated at the free surface of the fluid when
the forcing acceleration $\Gamma$ is increased above a threshold
$\Gamma_{0}$, which is a function of the forcing frequency $f$.
The waves are subharmonic, i.e. they oscillate at half the forcing
frequency, $f/2$. The wavelength $\lambda$ and wavenumber $k=2\upi/\lambda$
are related to the natural frequencies of the surface $f_{m,n}=\omega_{m,n}/\left(2\upi\right)$
through the dispersion relation for closed basins 
\begin{equation}
\omega_{m,n}^{2}=gk\tanh kd.\label{eq:dispersion}
\end{equation}
The quantity $k$ is the modulus of the wavenumber $\mathbf{k}=\left(k_{x},k_{y}\right)=\left(m\upi/L_{x},n\upi/L_{y}\right)$
and $m,n\in\mathbb{Z}$ are the number of nodes in each direction
(for more details, cf. \citealp{1976JFM....75..419M}). For a forcing
frequency $f$ close to $2f_{m,n}$, the systems responds with a wavelength
given by (\ref{eq:dispersion}). The sign of the detuning $\Delta f=f/2-f_{m,n}$
defines if the onset of the Faraday waves is subcritical ($\Delta f<0$)
or supercritical ($\Delta f>0$), as shown by \citet{1990JFM...221..383D}.

\subsection{Observation of the streaming patterns\label{sub:Observations}}

\begin{figure}
\begin{centering}
\includegraphics[width=1\textwidth]{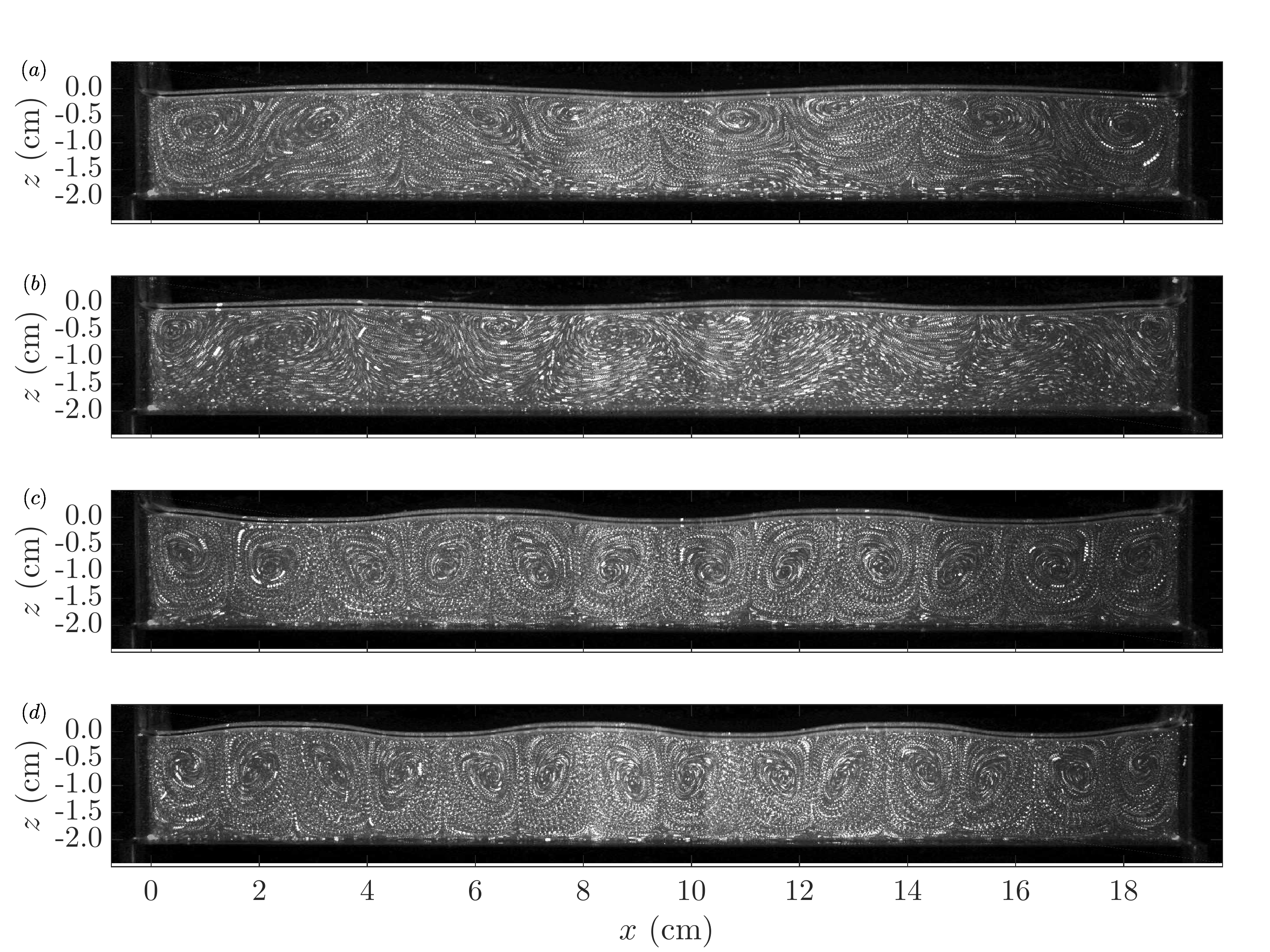}
\par\end{centering}

\protect\caption{Streaming patterns beneath purely longitudinal Faraday waves for (\emph{a})
$f=7.6\,\mathrm{Hz}$, (\emph{b}) $f=8.9\,\mathrm{Hz}$, (\emph{c})
$f=9.9\,\mathrm{Hz}$ and (\emph{e})\emph{ }$f=10.8\,\mathrm{Hz}$
in the midplane in the crosswise direction ($y=L_{y}/2$). The images
were obtained by strobe illumination. The number of nodes of the standing
Faraday wave is respectively $m=4,5,6,7$ ($n=0$). \label{fig:overexposed}}
\end{figure}

At first glance, the flow beneath the free surface is very simple:
the fluid is carried from one crest to the neighbouring ones and then
back through a cycle \citep{van1982album}. However, particles do
not come back exactly to the same position after each cycle. This
fact can be easily observed when taking images stroboscopically. An
overexposed image with strobe flashing, shows the slow tracks followed
by particles after several cycles (see figure~\ref{fig:overexposed}).
These tracks will be referred to as \emph{streaming patterns}, and
their study, as the streaming flow or velocity field related to it,
are the main goal of this work. 

Despite the streaming patterns were observed indirectly by \citet{1831RSPT..121..299F}
through the accumulation of particles on the bottom of the trough,
the first visualisation is due to \citet{1990JFM...221..383D}, who
used Kalliroscope particles for this purpose. 

The streaming patterns are generated by a secondary flow which is
very small compared to the main oscillatory flow. In our experiments,
the streaming velocity field is typically around 20 times weaker than
the oscillatory velocity field \citep{Gordillo:2014jf}. Despite its
weakness, the streaming patterns can be easily observed using a strobe
light as a consequence of the fact that the main flow (the dominant
velocity signal) is periodical whereas the streaming one is mainly
steady. Hence, the motion that particles undergo after a complete
cycle due to the oscillatory field is zero and only the contribution
of the streaming velocity field is observed. Notice that particles
under the free surface of a standing wave, which is the case of the
Faraday waves, have a Stokes drift \citep{Stokes:1847uv} equal to
zero after a complete cycle \citep{Gordillo:2012wi}.

The streaming patterns show also a weak dependence on the phase at
which the images are captured. Naturally, the patterns are distorted
with the deformation of the free surface at a given phase. For the
sake of simplicity, all the quantitative measurements of the streaming
field were performed at the phase at which the free surface is flat
and only on longitudinal Faraday waves (no crosswise component, i.e.
$n=0$ in equation \ref{eq:dispersion}). For the same reason, all
the measurements showed here are restrained to the midplane in the
$y$ direction.

The images in figure~\ref{fig:overexposed} also show that the structure
of the streaming patterns is highly three-dimensional, even for Faraday
waves with no crosswise component on the surface. Particles may enter
and exit the illuminated plane, which explains the observed spiral
trajectories followed by the particles.

\subsection{Measurement protocol\label{sub:Measurement-protocol}}

To characterise the dependence of the streaming patterns on the amplitude
and the frequency, we performed several experimental runs. The protocol
was automatised using Matlab, which controlled, via several interfaces
(GPIB, USB, RS232 and Ethernet) all the devices involved in the experiment.
Each set of measurements analysed a single forcing frequency. The
protocol was the following: First, the amplitude was increased until
stationary Faraday waves with large-amplitude emerge. A downward-ramp
scheme in amplitude was then followed: For each amplitude, we record
a first series of 36 images, which completed a cycle, to determine
the shape of the interface in terms of the phase (or time). Immediately
after, the shape of the free surface was detected using a standard
algorithm for edge detection. After the flat-state phase was determined,
the synchronisation signal was shifted and a second run consisting
of 64 stroboscopic flat-state images was recorded. Measurements were
done cyclically every four minutes, following the downward ramp until
the Faraday waves on the surface vanished.

In Table~\ref{tab:Measurements}, we display a list of the experiments
performed. The runs include different frequencies and different wavenumbers.
We also studied different nonlinear scenarios, i.e. subcritical and
supercritical behaviour. Measurement ramps lasted between 1 to 4 hours,
depending on the initial amplitude and the ramp step size. 

\begin{table}
\begin{centering}
\begin{tabular}{>{\centering}b{0.18\columnwidth}>{\centering}b{0.18\columnwidth}>{\centering}b{0.18\columnwidth}>{\centering}b{0.18\columnwidth}>{\centering}b{0.18\columnwidth}}
label &
$f\,\left(\mathrm{Hz}\right)$ &
number of nodes $\left(m\right)$ &
nonlinear

behaviour &
observed 

types\tabularnewline\addlinespace[2mm]
a &
7.6 &
4 &
subcritical &
II\tabularnewline
b &
8.9 &
5 &
subcritical &
III\tabularnewline
c &
9.9 &
6 &
subcritical &
I, III\tabularnewline
d &
9.95 &
6 &
supercritical &
I, III\tabularnewline
e &
10.8 &
7 &
subcritical &
I\tabularnewline
f &
10.85 &
7 &
supercritical &
I\tabularnewline
g,h,i,j &
10.8 &
7 &
subcritical &
I, II, III\tabularnewline
\end{tabular}
\par\end{centering}

\protect\caption{Set of measurements for the streaming patterns. Each line corresponds
to a ramp in amplitude. The set includes four different wavelengths.
Faraday waves destabilise through subcritical or supercritical bifurcations
depending on the frequency detuning. Both subcritical and supercritical
behaviours are explored. The last column shows the observed type of
streaming pattern. \label{tab:Measurements}}
\end{table}

\subsection{Experimental results\label{sub:Experimental-results}}

Our sequences of images were processed by means of two different techniques.
In a very simple way, the sequences can be used to detect the trajectories
followed by the seed particles. A synthetical over-exposed image is
obtained by maximising the grey value of each pixel along several
frames. This is actually how the images in figure~\ref{fig:overexposed}
were obtained. This strategy allows to fast probe the streaming patterns
and provides a powerful qualitative tool to analyse the data on the
run. The second technique is PIV, which provides precise quantitative
data. For this purpose, each sequence of images was analysed using
our own PIV code. After preparing masks and background subtraction,
we used time-averaging in the correlation-function space to improve
the signal-to-noise ratio \citep{Meinhart:2000wc}. This allows us
to obtain neat velocity fields with a high spatial resolution ($8\times8$-pixel
window size, equivalent to $0.64\times0.64\,\mathrm{mm^{2}}$ resolution)
and no spatial averaging (0\% overlapping).

\begin{figure}
\begin{centering}
\includegraphics[width=1\textwidth]{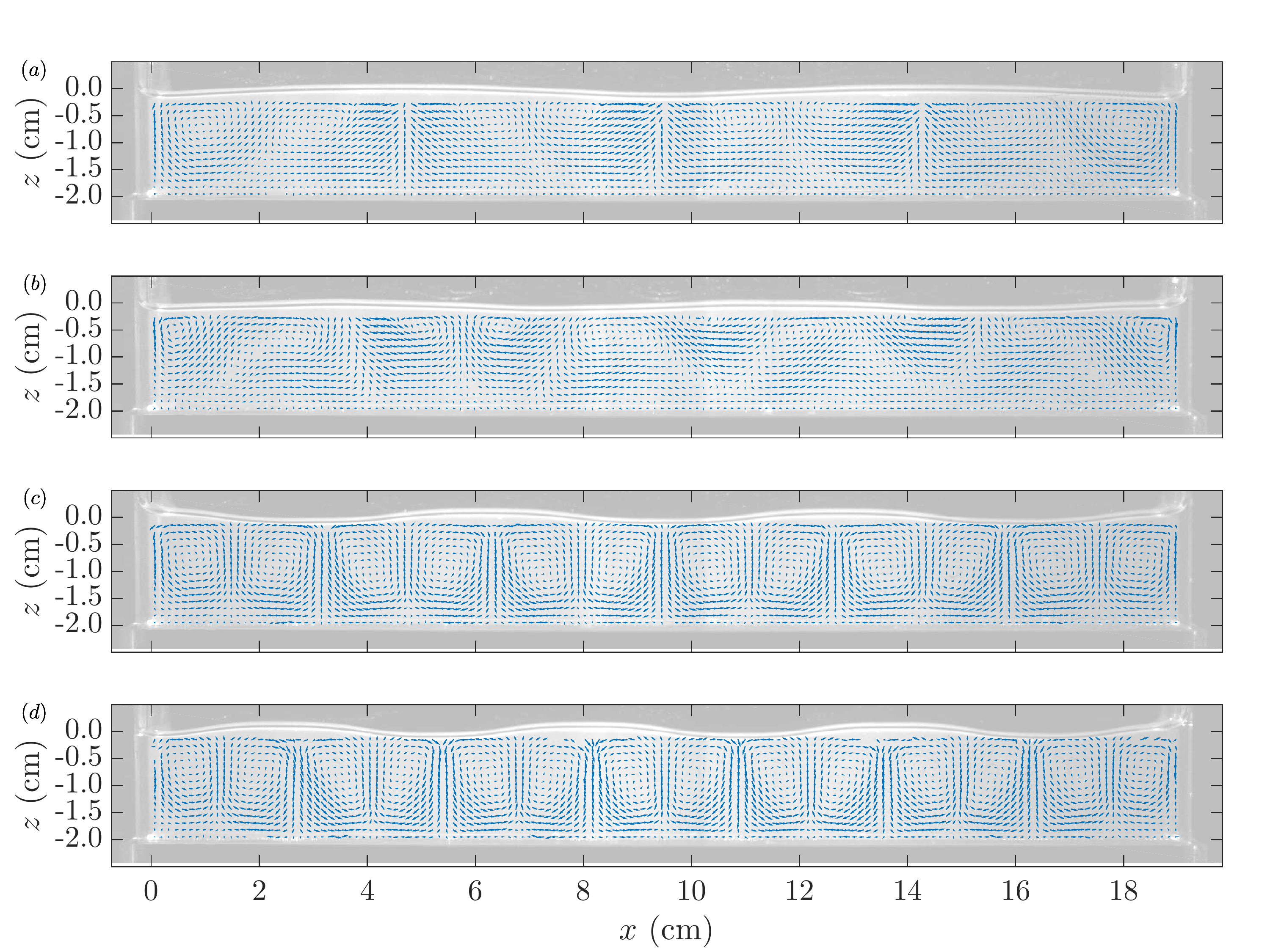}.
\par\end{centering}

\protect\caption{Velocity field of the streaming patterns shown in figure~\ref{fig:overexposed}
(increasing wavenumber from top to bottom). The background shows the
shape of the interface $20\,\mbox{ms}$ after the flat phase. Insets
(\emph{a}), (\emph{c}) and (\emph{d}) show ordered patterns with a
spatial period equal to the half of the surface waves. Inset (\emph{b})
shows a distorted streaming pattern. \label{fig:piv}}
\end{figure}

Figures~\ref{fig:overexposed} and \ref{fig:piv} show respectively
the streaming pattern and the velocity field $\left(u,w\right)$ in
the $\left(x,z\right)$-plane for four different forcing frequencies.
The presented cases are a good representation of the whole set of
collected data. In all our measurements, we observe: (i) a sequence
of counter-rotating structures with a wavelength equal to half the
free surface wavelength, (ii) that velocity field strength decreases
with depth, and (iii) the fluid is pushed downward at the free-surface
antinodes and upward at the nodes. However there are clear differences
between the flow structures in the subfigures presented in figures~\ref{fig:overexposed}
or \ref{fig:piv}. Based on our results and qualitative criteria,
we identify three types of streaming patterns: 
\begin{lyxlist}{00.00.0000}
\item [{Type~I:}] sequence of counter-rotating rolls (inset c-d); 
\item [{Type~II:}] sequence of moustache-shaped swirls (inset a); 
\item [{Type~III:}] irregular patterns (inset b).
\end{lyxlist}
Although only type I and II show ordered structures, type III displays
well-shaped rolls next to the lateral walls at $x=0$ and $x=L_{x}$.
Types I and II share another common feature: both display squeezed
counter-rotating structures at the bottom. 

To illustrate quantitatively the classification described previously,
we chose two quantities that characterise the state of the system.
The first one is the amplitude $\zeta$ of the Faraday wave and is
used to quantify the state of the free-surface. This quantity can
be obtained from the sequence of images used for determining the flat
state (see \S \ref{sub:Measurement-protocol}). As second parameter,
intended to represent the state of the flow, we chose the amplitude
(along the $x$ coordinate) of the vertically averaged velocity $w$,
i.e.
\[
\hat{w}\equiv\max_{x\epsilon D}\left|\frac{1}{d}\int_{-d}^{0}\mathrm{d}z\, w\left(x,y=b/2,z\right)\right|.
\]
Since the lateral walls induce strong coherent structures for all
the types, the domain $D$ is chosen such that the first and last
roll are removed, i.e. $D:\left[\lambda/4,L_{x}-\lambda/4\right]$.
The quantity $\hat{w}$ can be easily obtained from PIV data (the
discretised $\mathrm{d}z$ is $0.64\,\mathrm{mm}$). Averaging along
$z$ cancels noise induced by PIV detection. 

\begin{figure}
\begin{centering}
\includegraphics[scale=0.5]{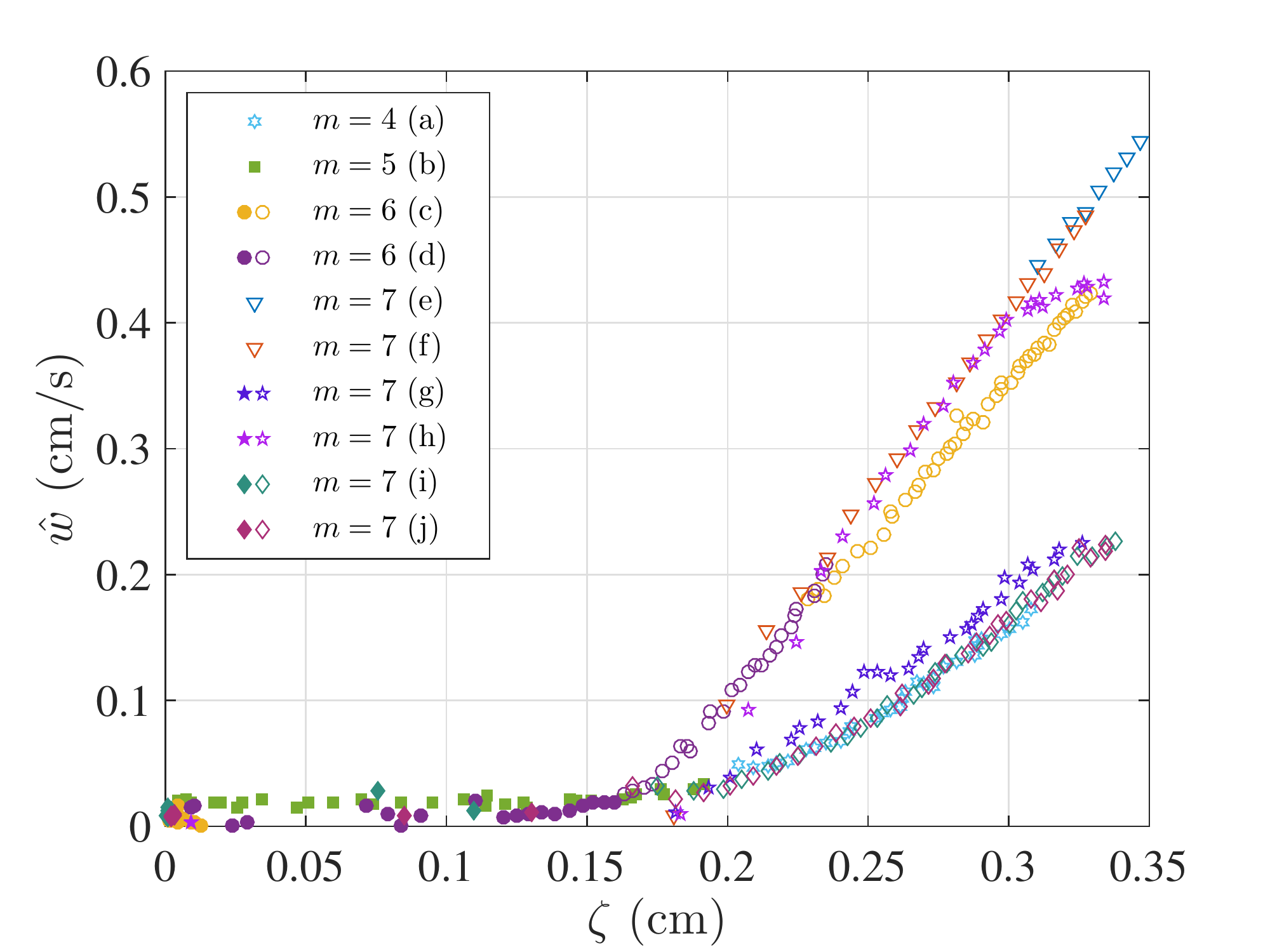}
\par\end{centering}

\protect\caption{Bifurcation diagram for streaming patterns. The diagram shows all
the collected data which collapse into two branches. The upper branch
contains Type I streaming patterns (roll structures) and the lower
one, Type II patterns. Type III (filled symbols) patterns can only
be observed for very low amplitudes. \label{fig:Bifurcation-diagram}}
\end{figure}

The two chosen quantities are plotted against each other in figure~\ref{fig:Bifurcation-diagram}
for all our measurements. For values of $\zeta$ below some threshold
$\zeta_{0}\approx\text{1.5\,\ mm}$, $\hat{w}$ values remains almost
equal to zero, which is related to the absence of periodic structures.
Beyond this threshold, $\hat{w}$ bifurcates into two branches, each
containing a large amount of collapsed measurement points. All type-I
streaming patterns collapse to the upper branch, while type-II patterns
collapse to the lower one. The upper branch displays $\hat{w}$-values
twice as strong as than the lower one. Type-III patterns can be observed
only for low $\hat{w}$ values, which gives a hint to understand their
existence. A reasonable explanation is that the coherent periodic
structures induced by the Faraday waves are so weak that they compete
with weak parasite large-scale eddies. The latter are presumably generated
by surface tension effects and persist for long times, even after
the vibration is switched off).

Figure~\ref{fig:Bifurcation-diagram} reveals that subcritical or
supercritical behaviour has not any relevant effect on streaming patterns:
series (c) and (d) form a continuous line for $m=6$; the same stands
for series (e) and (f), where $m=7$. More strikingly, Type I and
II patterns may emerge regardless the value of $\lambda$. The evidence
for $m=7$ is remarkable because both types are observed with the
same forcing parameters: type I, runs (e),(f) and (h) and type II,
runs (i) and (j). The mixture in the trough was the same and most
of the runs were taken during the same day. We could not find a way
to choose which streaming-flow pattern (I or II) was going to be selected
during an experiment: the system spontaneously chooses one and keeps
it until the wave on the surface vanishes. Multi-stability of viscous
modes have been observed in other flows, \emph{e.g.} \citet{1965JFM....21..385C}
observed more than 20 different states in the Taylor-Couette flow
under the same forcing parameters .

As a closing for this section, we would like to readdress the effect
of the streaming patterns at long timescales on particles in the liquid.
Consider a particle subjected to the flow under Faraday waves. Besides
the oscillating forces due to the fast oscillating velocity field,
our particle is subjected to a viscous steady force due to the streaming
flow. Hence, at long timescales, the particle will be slowly drifted
by this force until it reaches a stagnation point. Particles heavier
than the fluid will accumulate at stagnation points at the bottom
of the trough, creating the heaps observed by \citeauthor{1831RSPT..121..299F}
in his seminal experiments. Particles lighter than the fluid, will
float at the free surface, and accumulate into \emph{rafts }as has
been shown in experiments by \citet{Falkovich:2005ei} and \citet{Sanl:2014hs}.
The streaming flows are thus a relevant extra ingredient to understand
the motion induced on particles on Faraday waves. A step forward is
to build a theory that explains the emergence of the streaming patterns
observed in our experiments. The next section is devoted to this task.

\section{Theory\label{sec:Theory}}

We consider a three-dimensional infinite layer of an incompressible
and viscous fluid (density $\rho$ and kinematic viscosity $\nu$)
delimited at its bottom by an impermeable flat wall. The flow far
from the wall is tangent to the wall and periodic in time. Our purpose
is to determine the stationary component, or streaming, due to this
periodic flow. The streaming is the result of a complex mechanism
that couples the flows inside the bulk (far from the walls) and at
the boundary layers (near the walls). The coupling can be split in
three intermediate mechanisms:
\begin{enumerate}
\item the oscillatory flow in the bulk induces an oscillating boundary layer,
\item the oscillating boundary layer exerts a feedback on the bulk flow
whose steady component originates the streaming,
\item the streaming is diffused into the bulk due to a viscous process.
\end{enumerate}
In \S\ref{sub:streaming_general_theory} we present an extension
of Batchelor's two-dimensional model for oscillatory flows \citep[see][pp. 353-364]{Batchelor2000an_introduction}
to three dimensions whose basis is two-fold: the length scale separation
between the viscous boundary layer and the bulk flows; and the emergence
of steady terms due to nonlinear convective term at the boundary layer.
For the sake of clarity, we outline here our theoretical framework.
First, in \S\ref{S1}, we pose the generic problem of an oscillating
flow near a rigid wall. In \S\ref{S2} we determine the equations
in both the boundary layer and the bulk. Starting from the hypothesis
of weak quasi-inviscid waves, we define two small parameters: $\gamma,$
the ratio of the boundary layer thickness to the typical length scale
the bulk flow and $\beta,$ the ratio of the period of the oscillating
flow to the convective time scale. Navier-Stokes equations are accordingly
expanded in powers of $\gamma$ and $\beta$ inside the boundary layer
as well as in the bulk. Order by order, equations are truncated, solved
and matched at the junction. At dominant order (see \S\ref{sub:oscill-BL}),
an oscillatory boundary layer is obtained due to the oscillating field
at the bulk {[}mechanism (\emph{a}){]}. Higher-order equations in
$\beta$ and $\gamma$ yield an equation for the streaming flow inside
the boundary layer (see \S\ref{sub:Asymptotic-matching}). This streaming
flow induces a streaming velocity at the junction {[}mechanism (\emph{b}){]},
which can be identified as a \emph{matched} boundary condition at
the wall for the bulk streaming flow. Finally, we show in \S\ref{sub:Streaming-bulk}
that the streaming flow in the bulk can be computed by solving an
independent Navier-Stokes equations for the streaming velocities,
with the aforementioned matched boundary conditions. The result is
the diffusion of the streaming from the boundaries into the bulk {[}mechanism
(\emph{c}){]}. Finally, our theoretical framework is applied to find
the matched streaming boundary conditions for Faraday waves in \S\ref{S3},
which are required to solve the streaming Navier-Stokes equations
in $\S$\ref{sec:Numerical-simulations}.

\subsection{Three-dimensional streaming due to an oscillatory boundary layer
\label{sub:streaming_general_theory}}

\subsubsection{External oscillatory field\label{S1}}

\begin{figure}
\begin{centering}
\includegraphics[width=1\textwidth]{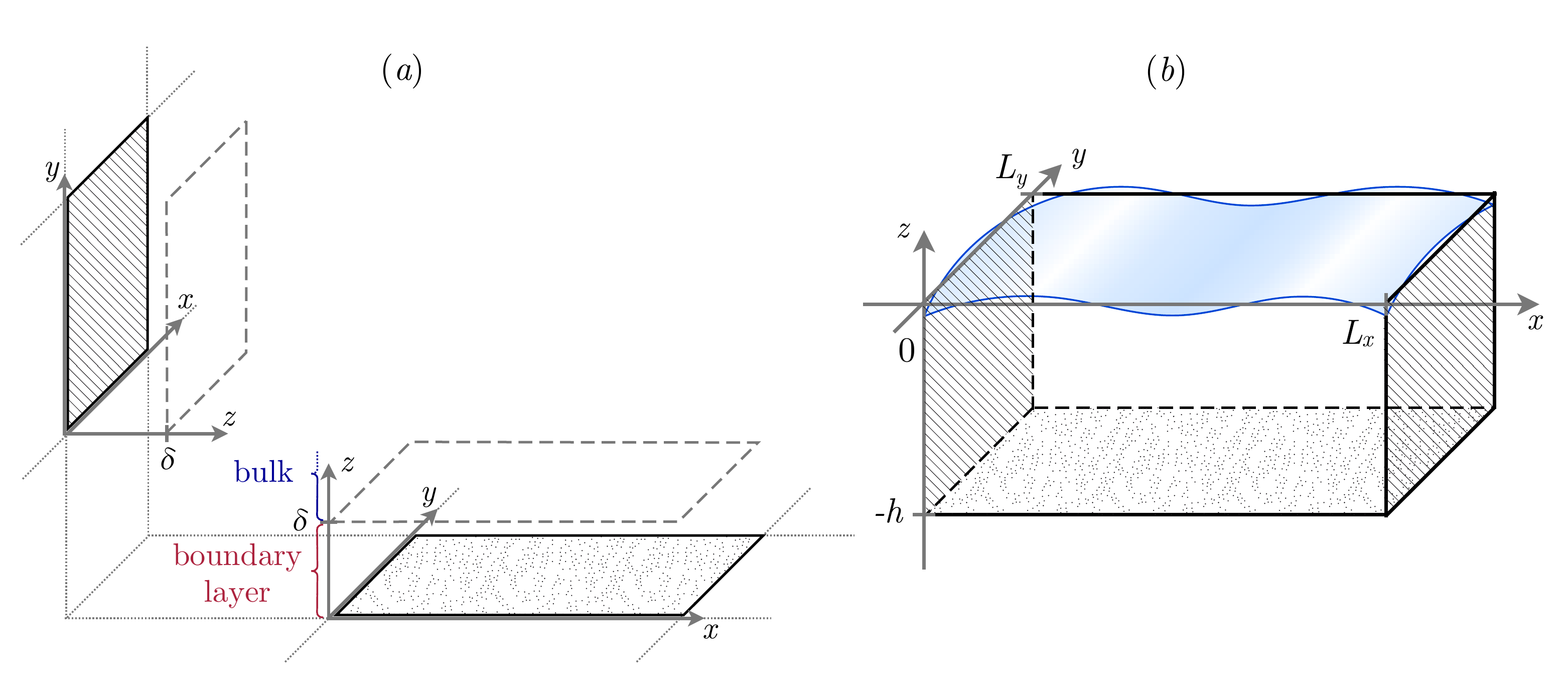}
\par\end{centering}

\centering{}\protect\caption{(\emph{a}) Local coordinate systems on two walls of the container.
The colored surfaces represent parts of the walls far from the edges
of the domain. The grey parallelograms show the limit of the boundary
layer at the distance $\delta$ from the walls. (\emph{b}) Definition
of the domain containing the fluid for numerical simulations\label{fig:bords}. }
\end{figure}

We denote a local coordinate system ($x,y,z$) as shown in figure~\ref{fig:bords}(\emph{a});
$u$, $v$ and $w$ are the components of the velocity along the $x$,
$y$ and $z$ directions, respectively. The local coordinates are
chosen such that the $z$ axis is normal to the wall and points towards
the bulk. The $z$-domain is split into two subdomains: an inner region
which comprises the boundary layer at the vicinity of the wall and
an outer region extending beyond the boundary layer. The fields inside
the boundary layer are denoted $u$, $v$, $w$ and $p$, standing
for the components of the velocity and the pressure. Outside the boundary
layer, the outer flow is denoted by the fields \uu, \uv, \uw\,
and \up. The outer flow is externally imposed, assumed to be tangent
at $z=0$, and given by the general form 
\begin{equation}
\begin{array}{c}
\vs\dps{(\uum,\uvm,\uwm)=\Real\left\{ \left(U_{0}\left(x,y,z\right),V_{0}\left(x,y,z\right),W_{0}\left(x,y,z\right)\right)\ex^{\ic\omega t}\right\} ,}\\
\left.\uwm\right|_{z=0}=0.
\end{array}\label{eq-fields_out}
\end{equation}
The length scale of variations in the three directions $x,y$ and
$z$ are assumed to be of order $L$. As we will see, this outer flow
induces inside the boundary layer a motion influenced by viscous stresses.

\subsubsection{Boundary layer equations \label{S2}}

We place ourselves inside the oscillating boundary layer and consider
the full Navier-Stokes equations for an incompressible flow: 
\begin{equation}
\begin{array}{c}
\vs\dps{\ddt u+u\ddx u+v\ddy u+w\ddz u=-\frac{1}{\rho}\ddx p+\nu\ddxx u+\nu\ddyy u+\nu\ddzz u},\\
\vs\dps{\ddt v+u\ddx v+v\ddy v+w\ddz v=-\frac{1}{\rho}\ddy p+\nu\ddxx v+\nu\ddyy v+\nu\ddzz v},\\
\vs\dps{\ddt w+u\ddx w+v\ddy w+w\ddz w=-\frac{1}{\rho}\ddz p+\nu\ddxx w+\nu\ddyy w+\nu\ddzz w},\\
\ddx u+\ddy v+\ddz w=0.
\end{array}\label{eq-NS2_in}
\end{equation}
We rewrite the involved physical fields in terms of dimensionless
quantities: 
\begin{equation}
\begin{array}{c}
\vs\dps{u=U_{0}u',\qquad v=U_{0}v',\qquad w=U_{0}w',\qquad p=\rho\omega LU_{0}p',}\\
\vs\dps{\ddx=\frac{1}{L}\partial_{x'},\qquad\ddy\sim\frac{1}{L}\partial_{y'},\qquad\ddz\sim\frac{1}{\delta}\partial_{z'},\qquad\ddt\sim\omega\partial_{t'}.}
\end{array}\label{Eq_char_amp}
\end{equation}
Here, the length scale $\delta$ is defined as $\sqrt{2\nu/\omega}$
and represents the extent of the boundary layer. Based on this, we
define two dimensionless parameters $\gamma$ and $\beta$: 
\begin{equation}
\gamma\equiv\frac{\delta}{L}\ll1,\qquad\beta\equiv\frac{U_{0}}{\omega L}\ll1.\label{eq:beta_gamma}
\end{equation}
The quantity $\gamma$ represents the ratio between the boundary layer
extent $\delta$ and the outer-flow length scale $L$ while $\beta$
is the ratio between the period of the oscillating flow and the convective
timescale (a low $\beta$ implies weak convection). Using the previous
definitions, the set of equations in (\ref{eq-NS2_in}) are hence
reduced to 
\begin{equation}
\begin{array}{c}
\vs\dps{\ddt\left(u,v\right)+\beta\left(u\ddx+v\ddy+\frac{1}{\gamma}w\ddz\right)\left(u,v\right)=-\left(\ddx,\ddy\right)p+\frac{1}{2}\left[\gamma^{2}\left(\ddxx+\ddyy\right)+\ddzz\right]\left(u,v\right)},\\
\gamma\vs\dps{\ddt w+\beta\gamma\left(u\ddx w+v\ddy w+\frac{1}{\gamma}w\ddz w\right)=-\ddz p+\frac{1}{2}\left[\gamma^{3}\left(\ddxx+\ddyy\right)+\gamma\ddzz\right]w},\\
\gamma\left(\ddx u+\ddy v\right)+\ddz w=0,
\end{array}\label{eq-NS2_in_expanded}
\end{equation}
where primes have been omitted for simplicity. We now expand the fields
in powers of $\beta$ and $\gamma$: 
\begin{equation}
\left[\begin{array}{c}
u\\
v\\
w\\
p
\end{array}\right]=\left[\begin{array}{c}
u_{0}\\
v_{0}\\
0\\
p_{0}
\end{array}\right]+\beta\left[\begin{array}{c}
u_{1}\\
v_{1}\\
w_{1}\\
p_{1}
\end{array}\right]+\gamma\left[\begin{array}{c}
\tilde{u}_{1}\\
\tilde{v}_{1}\\
\tilde{w}_{1}\\
\tilde{p}_{1}
\end{array}\right]+\ldots\label{eq:expansion}
\end{equation}
This expansion is plugged into (\ref{eq-NS2_in_expanded}), providing
a hierarchy of equations of powers of $\beta$ and $\gamma$%
\footnote{Since $\beta$ and $\gamma$ have a priori different and unknown orders
of magnitude, identifications as $\beta^{p}=\gamma^{q}$ or subsequent
simplifications are not consistent.%
}. The first equations of the hierarchy are: 
\begin{equation}
\mathcal{O}\left(1\right):\:\:\left\{ \begin{array}{l}
\vs\dps{\ddt\left(u_{0},v_{0}\right)=-\left(\ddx,\ddy\right)p_{0}+\frac{1}{2}\ddzz\left(u_{0},v_{0}\right)},\\
\ddz p_{0}=0
\end{array}\right.\label{eq-exp_1}
\end{equation}

\begin{equation}
\mathcal{O}\left(\gamma\right):\:\:\left\{ \begin{array}{l}
\vs\dps{\ddz\tilde{p}_{1}=0},\\
\ddx u_{0}+\ddy v_{0}+\ddz\tilde{w}_{1}=0
\end{array}\right.\label{eq-exp_gamma}
\end{equation}
\begin{equation}
\mathcal{O}\left(\beta\right):\:\:\left\{ \begin{array}{l}
\vs\dps{\ddt\left(u_{1},v_{1}\right)=-\left(u_{0}\ddx+v_{0}\ddy+\tilde{w}_{1}\ddz\right)\left(u_{0},v_{0}\right)-\left(\ddx,\ddy\right)p_{1}+\frac{1}{2}\ddzz\left(u_{1},v_{1}\right)},\\
\vs\ddz p_{1}=0,\\
\ddz w_{1}=0.
\end{array}\right.\label{eq-exp_beta}
\end{equation}

A similar analysis can be performed for the outer flow. In that case,
the rescaling rules from (\ref{Eq_char_amp}) apply except for $\ddz=\frac{1}{\delta}\partial_{z'}$,
which should be replaced by $\ddz\approx\frac{1}{L}\partial_{z'}$.
Likewise, a zero-th order term $\uwm_{0}$ should be included in the
expansion in powers of $\beta$ and $\gamma$ (\ref{eq:expansion})%
\footnote{The normal velocity condition $\uwm_{0}=0$ is required only at $z=0$. %
}. Order by order, it follows that 
\begin{equation}
\mathcal{O}\left(1\right):\:\:\left\{ \begin{array}{l}
\vs\dps{\ddt\left(\uum_{0},\uvm_{0},\uwm_{0}\right)=-\left(\ddx,\ddy,\ddz\right)\upm_{0}},\\
\ddx\uum_{0}+\ddy\uvm_{0}+\ddz\uwm_{0}=0,
\end{array}\right.\label{eq-out_exp_1}
\end{equation}
\begin{equation}
\mathcal{O}\left(\beta\right):\:\:\begin{array}{l}
\dps{\ddt\left(\uum_{1},\uvm_{1},\uwm_{1}\right)=\left(-\uum_{0}\ddx-\uvm_{0}\ddy-\uwm_{0}\ddz\right)\left(\uum_{0},\uvm_{0},\uwm_{0}\right)-\left(\ddx,\ddy,\ddz\right)\upm_{1}}.\end{array}\label{eq-out_exp_beta}
\end{equation}
To find asymptotically uniform solutions in the whole domain, the
final step is to match the fields at the junction \citep{Bender1999advanced}.
From a mathematical point of view, this is equivalent to impose that
the fields in the inner flow $\left(u,v,w,p\right)$ evaluated in
the limit $z\text{'\ensuremath{\rightarrow\infty}}$ are equal to
those in the outer flow, $\left(\uum_{1},\uvm_{1},\uwm_{1},\upm_{1}\right)$
evaluated in limit $z\rightarrow0$.

\subsubsection{Zeroth order: The oscillatory boundary layer\label{sub:oscill-BL}}

The equations for the dominant variables $u_{0},v_{0}$ and $p_{0}$
are given by (\ref{eq-exp_1}). Notice that including a $w_{0}$ term
in the expansion (\ref{eq:expansion}), yields $\partial_{z}w_{0}=0$
at leading order, which requires $w_{0}=0$ to satisfy the non-slip
boundary condition at $z=0$. On the other hand, the $z$-independence
of $p_{0}$ implies that the pressure is constant across the boundary
layer. The matching condition for the pressure becomes trivial and
(\ref{eq-out_exp_1}) can be used to eliminate the pressure from the
unknowns. Accordingly, $u_{0}$ relates to $\uum_{0}$ as follows
\begin{equation}
\begin{array}{c}
\vs\dps{\ddt\left(u_{0},v_{0}\right)=\vs\ddt\left(\uum_{0},\uvm_{0}\right)+\nu\ddzz\left(u_{0},v_{0}\right)}.\end{array}\label{eq-NS1_inout-1}
\end{equation}
The equation (\ref{eq-NS1_inout-1}) with the prescribed boundary
conditions admits the solution 
\begin{equation}
\begin{array}{c}
\vs\dps{u_{0}\left(x,y,z,t\right)=\Real\left\{ U_{0}\left(x,y\right)\ex^{\ic\omega t}\left(1-\ex^{-\alpha z}\right)\right\} },\\
\dps{v_{0}\left(x,y,z,t\right)=\Real\left\{ V_{0}\left(x,y\right)\ex^{\ic\omega t}\left(1-\ex^{-\alpha z}\right)\right\} },
\end{array}\label{eq-fields_in-1}
\end{equation}
where $\alpha=\left(1+\ic\right).$ Notice that the boundary layer,
in this approximation oscillates at the same frequency as the bulk
fields. In the case $U_{0}\left(x,y,z\right)=U_{0},$ $V_{0}\left(x,y,z\right)=V_{0}$
and $W_{0}\left(x,y,z\right)=0$, this solution reduces to the oscillatory
boundary layer found by \citet{1851TCaPS...9....8S}, which are an
exact solution of the Navier-Stokes equation.

\subsubsection{The induced streaming\label{sub:Asymptotic-matching}}

From the equations (\ref{eq-exp_beta}), it can be easily shown that
$p_{1}$ and $w_{1}$ follow the same rules as $p_{0}$ and $w_{0}$.
Up to this point, we notice that the first non-trivial term of the
expansion for $w$ is proportional to $\gamma$, which means that
$w$ scales as $\delta U_{0}/L$ and not as $U_{0}$. Remarkably,
although at zeroth order $w$ vanishes, at higher order, $w$ induces
a flow inside the boundary layer as will be shown in the following
calculations. From the divergence equation in (\ref{eq-exp_gamma})
and the variables $u_{0}$ and $v_{0}$ computed in (\ref{eq-fields_in-1}),
we deduce $\tilde{w}_{1}$, whose expression, 
\begin{equation}
\tilde{w}_{1}\left(x,y,z,t\right)=-\int_{0}^{z}\Real\left\{ \left(\ddx U_{0}+\ddy V_{0}\right)\ex^{\ic\omega t}\left(1-\ex^{-\alpha z}\right)\right\} \,\dd z,\label{eq-wtilde1}
\end{equation}
consistently satisfies the no-slip boundary condition at the wall
$\tilde{w}_{1}=0$. Since $p_{1}$ is also constant across the boundary
layer, $p_{1}$ equals ${\upm}_{1}$ all along it, which is used to
link the first equation in (\ref{eq-exp_beta}) and \ref{eq-out_exp_beta}:
\begin{equation}
\begin{array}{c}
\vs\ddt u_{1}+u_{0}\ddx u_{0}+v_{0}\ddy u_{0}+\tilde{w}_{1}\ddz u_{0}=\ddt\uum_{1}+\uum_{0}\ddx\uum_{0}+\uvm_{0}\ddy\uum_{0}+\frac{1}{2}\ddzz u_{1},\\
\vs\ddt v_{1}+u_{0}\ddx v_{0}+v_{0}\ddy v_{0}+\tilde{w}_{1}\ddz v_{0}=\ddt\uvm_{1}+\uum_{0}\ddx\uvm_{0}+\uvm_{0}\ddy\uvm_{0}+\frac{1}{2}\ddzz v_{1},\\
w_{1}=0.
\end{array}\label{eq-uvwfc}
\end{equation}
The streaming is extracted from (\ref{eq-uvwfc}), by averaging over
one period of oscillation $T=2\pi/\omega$. The nonlinear terms provide
components that oscillate at the frequency $2\omega$ and some that
are steady. To match the nonlinear terms of (\ref{eq-uvwfc}), $u_{1}$,
$v_{1}$, $\uum_{1}$ and $\uvm_{1}$ are assumed to have the generic
form 
\begin{equation}
\mathcal{A}\left(x,y,z\right)\ex^{\ic2\omega t}+\mathcal{B}\left(x,y,z\right)+\mathcal{A}^{*}\left(x,y,z\right)\ex^{-\ic2\omega t},\label{eq-u1_v1}
\end{equation}
 where $^{*}$ denotes the complex conjugate operator, $\mathcal{A}$
and $\mathcal{A}^{*}$ are the oscillating components of the fields
and $\mathcal{B}$ is stationary and responsible for the streaming
in the bulk. Then we focus on $\mathcal{B}$. We define the temporal
average $\langle f(t)\rangle$ of a function $f(t)$ as $\langle f\left(t\right)\rangle=\frac{1}{T}\int_{0}^{T}f\left(t\right)\dd t,$
where $T=2\upi/\omega$. Taking the generic form of $u_{1}$ (\ref{eq-u1_v1}),
we can easily deduce that $\langle\ddt u_{1}\rangle=0$ as well as
$v_{1}$, $\uum_{1}$ and $\uvm_{1}$. The system (\ref{eq-uvwfc})
then simplifies to a system of ordinary differential equations (ODEs)
in $\langle u_{1}\rangle$, $\langle v_{1}\rangle$ and $\langle w_{1}\rangle$
after averaging: 
\begin{equation}
\begin{array}{c}
\vs\frac{1}{2}\ddzz\langle u_{1}\rangle=\langle u_{0}\ddx u_{0}\rangle+\langle v_{0}\ddy u_{0}\rangle+\langle\tilde{w}_{1}\ddz u_{0}\rangle-\langle\uum_{0}\ddx\uum_{0}\rangle-\langle\uvm_{0}\ddy\uum_{0}\rangle,\\
\vs\frac{1}{2}\ddzz\langle v_{1}\rangle=\langle u_{0}\ddx v_{0}\rangle+\langle v_{0}\ddy v_{0}\rangle+\langle\tilde{w}_{1}\ddz v_{0}\rangle-\langle\uum_{0}\ddx\uvm_{0}\rangle-\langle\uvm_{0}\ddy\uvm_{0}\rangle,\\
\langle w_{1}\rangle=0.
\end{array}\label{eq-uvwfc_averaged}
\end{equation}
Starting from here we focus on $\langle u_{1}\rangle$ only, since
$\langle v_{1}\rangle$ can be found by analogy. The temporal averages
of each term in (\ref{eq-uvwfc_averaged}) are given below: 
\begin{equation}
\begin{array}{c}
\vs\dps{\langle u_{0}\ddx u_{0}\rangle=\frac{1}{4}\left(1-\ex^{-\alpha z}\right)\left(1-\ex^{-\alpha^{*}z}\right)\ddx\left(U_{0}U_{0}^{*}\right)},\\
\vs\dps{\langle v_{0}\ddy u_{0}\rangle=\frac{1}{4}\left(1-\ex^{-\alpha z}\right)\left(1-\ex^{-\alpha^{*}z}\right)\left(V_{0}\ddy U_{0}^{*}+V_{0}^{*}\ddy U_{0}\right)},\\
\vs\dps{\langle\tilde{w}_{1}\ddz u_{0}\rangle=-\frac{1}{2}\Real\left\{ \left(\ddx U_{0}+\ddy V_{0}\right)U_{0}^{*}\left[\left(z\alpha^{*}-\frac{\alpha^{*}}{\alpha}\right)\ex^{-\alpha^{*}z}+\frac{\alpha^{*}}{\alpha}\ex^{-(\alpha+\alpha^{*})z}\right]\right\} },\\
\vs\dps{\langle\uum_{0}\ddx\uum_{0}\rangle=\frac{1}{4}\ddx(U_{0}U_{0}^{*})},\\
\dps{\langle\uvm_{0}\ddy\uum_{0}\rangle=\frac{1}{4}\left(V_{0}\ddy U_{0}^{*}+V_{0}^{*}\ddy U_{0}\right)}.
\end{array}
\end{equation}
Summing all these terms together we establish the ODE that satisfies
$\langle u_{1}\rangle$ 
\begin{equation}
\begin{array}{rl}
\vs\dps{\ddzz\langle u_{1}\rangle}= & \dps{G_{1}\left(z\right)\Real\left(U_{0}\ddx U_{0}^{*}+V_{0}\ddy U_{0}^{*}\right)+\Real\left[G_{2}\left(z\right)(\ddx U_{0}+\ddy V_{0})U_{0}^{*}\right],}\end{array}\label{eq-dzz_u_averaged}
\end{equation}
where
\begin{equation}
\begin{array}{c}
G_{1}\left(z\right)=(1-\ex^{-\alpha z})(1-\ex^{-\alpha^{*}z})-1,\\
G_{2}\left(z\right)=-\left[\left(z\alpha^{*}-\frac{\alpha^{*}}{\alpha}\right)\ex^{-\alpha^{*}z}+\frac{\alpha^{*}}{\alpha}\ex^{-(\alpha+\alpha^{*})z}\right].
\end{array}\label{eq:G1_G2}
\end{equation}
For simplicity in what follows, the right-hand side of (\ref{eq-dzz_u_averaged})
is denoted $G(x,y,z)$. The associated boundary conditions are 
\begin{equation}
\begin{array}{c}
\vs\dps{\left.\langle u_{1}\rangle\right|_{z=0}=0},\\
\dps{\ddz\left.\langle u_{1}\rangle\right|_{z=\infty}=0},
\end{array}\label{eq-BCu_1averaged}
\end{equation}
from which we compute the streaming component: 
\begin{equation}
\langle u_{1}\rangle=\int_{0}^{z}\int_{\infty}^{z'}G(x,y,z'')\dd z''\dd z'.\label{eq-edo_u_1averaged}
\end{equation}
At the junction,\emph{ i.e.} in the limit $z\rightarrow\infty$, $\langle u_{1}\rangle$
must match $\langle\uum_{1}\rangle$ in the limit $z\rightarrow0$,
hence 
\begin{equation}
\begin{array}{rl}
\lim_{z\rightarrow0}\vs\langle\uum_{1}\rangle=\dps{\lim_{z\rightarrow\infty}\langle u_{1}\rangle}= & -\dps{\left[\frac{1}{(\alpha+\alpha^{*})^{2}}-\frac{\alpha^{2}+\alpha^{*2}}{\alpha^{2}\alpha^{*2}}\right]\Real\left(U_{0}\ddx U_{0}^{*}+V_{0}\ddy U_{0}^{*}\right)}\\
 & \dps{+\Real\left\{ \left[\frac{2}{\alpha^{*2}}-\frac{1}{\alpha\alpha^{*}}+\frac{\alpha^{*}}{\alpha(\alpha+\alpha^{*})^{2}}\right](\ddx U_{0}+\ddy V_{0})U_{0}^{*}\right\} }.
\end{array}\label{eq-u_1averaged_lim}
\end{equation}
The latter equation can be replaced in (\ref{eq:expansion}) to obtain
$\left(u,v,w\right)$, which in turn can be rescaled back to the original
physical variables via (\ref{eq:beta_gamma}) and (\ref{Eq_char_amp}).
This yields 
\begin{equation}
\begin{array}{rl}
\vs\lim_{z\rightarrow0}\langle\uum\rangle= & \dps{-\frac{1}{4\omega}\left[\Real\left(U_{0}\ddx U_{0}^{*}+V_{0}\ddy U_{0}^{*}\right)\right]}\\
 & +\dps{\frac{1}{8\omega}\left[(3\ic-2)(\ddx U_{0}+\ddy V_{0})U_{0}^{*}-(3\ic+2)(\ddx U_{0}^{*}+\ddy V_{0}^{*})U_{0}\right]}.
\end{array}\label{eq-streaming_u1_dim}
\end{equation}
It must be noted that the analysis at order $\gamma$ is similar to
that of (\ref{eq-exp_1}) and yields oscillating solutions which cancel
when temporally averaged. Hence, only the perturbation at order $\beta$,
which contains nonlinearities, generates the streaming. In other words,
the streaming emerges from the nonlinear nature of the convective
term of Navier-Stokes equation near the rigid wall.\\
\\
 The $v$ component of the streaming can be calculated straightforwardly
by applying the permutations $U_{0}\leftrightarrow V_{0}$ and $\ddx\leftrightarrow\ddy$.
This leads to 
\begin{equation}
\begin{array}{rl}
\vs\lim_{z\rightarrow0}\langle\uvm\rangle= & \dps{-\frac{1}{4\omega}\left[\Real\left(U_{0}\ddx V_{0}^{*}+V_{0}\ddy V_{0}^{*}\right)\right]}\\
 & +\dps{\frac{1}{8\omega}\left[(3\ic-2)(\ddx U_{0}+\ddy V_{0})V_{0}^{*}-(3\ic+2)(\ddx U_{0}^{*}+\ddy V_{0}^{*})V_{0}\right]}.
\end{array}\label{eq-streaming_v1_dim}
\end{equation}
The equations (\ref{eq-streaming_u1_dim}) and (\ref{eq-streaming_v1_dim})
describe the streaming at the junction of the bulk and the boundary
layer. In vector notation, we rewrite the latter expresion as: 

\begin{equation}
\begin{array}{rl}
\lim_{z\rightarrow0}\langle\underline{\mathbf{u}}\rangle & =\dps{-\frac{1}{4\omega}}\left\{ \Real\left[\left(\textbf{U}_{0}^{*}\cdot\nabla\right)\textbf{U}_{0}+2\textbf{U}_{0}^{*}\nabla\cdot\textbf{U}_{0}\right]+3\Imag\left(\textbf{U}_{0}^{*}\nabla\cdot\textbf{U}_{0}\right)\right\} \end{array}\label{eq-streaming_u1_vect}
\end{equation}
where $\textbf{U}_{0}=(U_{0},V_{0},0)$ and the component of $\textbf{U}_{0}$
normal to the wall is uniformly 0 so neither $w$ nor any of its derivatives
appear in (\ref{eq-streaming_u1_vect}). \\
A simpler expression can be obtained for flows that oscillate at the
same phase in both directions, \emph{i.e.} $\textbf{U}_{0}=\bm{\mathcal{U}_{0}}(x,y)\ex^{\ic\Theta(x,y)}$,
where the magnitude $\bm{\mathcal{U}_{0}}$ is a real vector (with
zero $z$-component) and the phase $\Theta$ is a real scalar: 
\begin{equation}
\lim_{z\rightarrow0}\langle\underline{\mathbf{u}}\rangle=-\frac{1}{4\omega}\left[\left(\bm{\mathcal{U}_{0}}\cdot\nabla\right)\bm{\mathcal{U}_{0}}+2\,\bm{\mathcal{U}_{0}}\nabla\cdot\bm{\mathcal{U}_{0}}+3\,\bm{\mathcal{U}_{0}}\left(\bm{\mathcal{U}_{0}}\cdot\nabla\right)\Theta\right].\label{eq-streaming_u1_vect2}
\end{equation}
The case of the two-dimensional streaming flow found by \citet{Batchelor2000an_introduction}
is 
\begin{equation}
\lim_{z\rightarrow0}\langle\underline{u}\rangle=-\frac{3}{4\omega}\left(\mathcal{U}_{0}\partial_{x}\mathcal{U}_{0}+\mathcal{U}_{0}^{2}\partial_{x}\Theta\right),\label{eq-streaming_u1_vect2_2D}
\end{equation}
which consistently matches the three-dimensional one, (\ref{eq-streaming_u1_vect2}),
when one of the two components of $\textbf{U}_{0}$ is imposed to
be 0.

\subsection{Streaming in the bulk\label{sub:Streaming-bulk}}

The final step is to determine how the streaming due to the boundary
layers induces a streaming in the bulk. For this purpose, consider
the general Navier-Stokes equation for the bulk in a comoving frame
of reference,
\begin{equation}
\partial_{t}\underline{\mathbf{u}}+\left(\underline{\mathbf{u}}\cdot\nabla\right)\underline{\mathbf{u}}=-\frac{\text{1}}{\rho}\nabla\underline{p}+a\left(t\right)\hat{\mathbf{k}}+\nu\nabla^{2}\underline{\mathbf{u}},\label{eq:NS-bulk}
\end{equation}
where $a\left(t\right)=-g+\Gamma\cos\omega t$ is the vertical acceleration
of the system.%
\footnote{ Capillary terms can also be included. %
} To simplify the notation, we omit the underline. We decompose the
velocity and pressure field into oscillatory and steady terms: $\mathbf{u}\left(\mathbf{x},t\right)=\widetilde{\mathbf{u}}\left(\mathbf{x},t\right)+\overline{\mathbf{u}}\left(\mathbf{x}\right)$
and $p\left(\mathbf{x},t\right)=\widetilde{p}\left(\mathbf{x},t\right)+\overline{p}\left(\mathbf{x}\right)$.
For most Faraday waves experiments, since $\lambda,h,L_{x},L_{y}\gg\delta$
the flow under the waves is potential at leading order, i.e. $\widetilde{\bm{\textbf{\ensuremath{\omega}}}}=\nabla\times\widetilde{\mathbf{u}}\left(\mathbf{x},t\right)=0$,
where $\bm{\omega}$ is the vorticity. Linear and nonlinear solutions
for the oscillatory part have been widely analyzed in the literature
\citep[see][ and references therein]{1993JFM...248..671M}. 

The convective term in (\ref{eq:NS-bulk}) can be written in terms
of $\bm{\omega}$. It can be easily shown that $\left(\mathbf{u}\cdot\nabla\right)\mathbf{u}=\nabla\left(\frac{1}{2}\widetilde{u}^{2}+\frac{1}{2}\overline{\mathbf{u}}^{2}+\widetilde{\mathbf{u}}\text{\ensuremath{\cdot}}\overline{\mathbf{u}}\right)-\left(\widetilde{\mathbf{u}}+\overline{\mathbf{u}}\right)\times\overline{\bm{\omega}}$.
Time averaging of the latter expression yields $\left\langle \left(\mathbf{u}\cdot\nabla\right)\mathbf{u}\right\rangle =\nabla\left\langle \frac{1}{2}\widetilde{u}^{2}\right\rangle +\left(\overline{\mathbf{u}}\cdot\nabla\right)\overline{\mathbf{u}}$.
Likewise, time averaging of (\ref{eq:NS-bulk}) leads to
\begin{equation}
\partial_{t}\overline{\mathbf{u}}+\left(\overline{\mathbf{u}}\cdot\nabla\right)\overline{\mathbf{u}}=-\frac{\text{1}}{\rho}\nabla\left(\overline{p}+\rho gz+\frac{1}{2}\rho\left\langle \tilde{u}^{2}\right\rangle \right)+\nu\nabla^{2}\overline{\mathbf{u}}.\label{eq:NS-bulk-streaming}
\end{equation}
By defining $p'\equiv\overline{p}+\rho gz+\frac{1}{2}\rho\left\langle \tilde{u}^{2}\right\rangle $,
the equation for the evolution of the steady part of $\mathbf{u}$
becomes identical to a Navier-Stokes equation with no external forcing.
It can also be shown that the incompressibility condition for the
steady field remains unchanged so $\nabla\cdot\overline{\mathbf{u}}=0$.
The boundary conditions to solve the system are those provided in
the junction between the boundary layers and the bulk found in \S\ref{sub:streaming_general_theory},
\emph{e.g.} (\ref{eq-streaming_u1_vect2}). This will be referred
to as the matched boundary condition. Thus, the streaming flow generated
in the bulk can be computed by solving classic Navier-Stokes equations
with non-trivial boundary conditions.

Remarkably, the contribution due to the local oscillatory velocity
field only appears in the effective pressure in (\ref{eq:NS-bulk-streaming})
and has null effect on $\overline{{\bf u}}$ at leading order. The
steady component of the flow stemming from the nonlinearities in the
bulk is thus negligible in comparison with the streaming generated
by the boundary layer. Henceforth, we expect that the boundary layers
are the main responsible for the appearance of streaming in quasi-inviscid
stationary waves.

\subsection{Classical Faraday waves\label{S3}}

\subsubsection{General case}

Here, we apply our previous findings to the classical Faraday waves.
The container is assumed to be a rectangular impermeable tank of horizontal
dimensions $L_{x}$, $L_{y}$ filled up to height $h$. The fluid
then occupies the domain $\left(x,y,z\right)\in\left[0,L_{x}\right]\times\left[0,L_{y}\right]\times\left[-h,0\right]$
as shown in figure~\ref{fig:bords}(\emph{b}). 

In the limit of ideal flows, the instantaneous fields in Faraday waves
are modelled by a velocity potential $\Phi$ inside the bulk, 
\begin{equation}
\Phi(x,y,z,t)=A\cos(k_{1}x)\cos(k_{2}y)\frac{\cosh\left(k(z+h)\right)}{\cosh(kh)}\cos(\omega t),\label{eq-potential}
\end{equation}
where $k_{1}$ and $k_{2}$ are multiples of $\upi/L_{x}$ and $\upi/L_{y}$
respectively, and $k=\sqrt{k_{1}^{2}+k_{2}^{2}}$. $\textbf{U}_{0}$
is related to $\Phi$ through $\textbf{U}_{0}=\nabla\Phi.$ Hence,
the oscillatory tangential velocity field at the boundaries can be
evaluated via equation (\ref{eq-potential}) and then plugged into
equation (\ref{eq-streaming_u1_vect2}) to obtain the matched boundary
condition for the streaming field. These streaming boundary conditions
will then be used in \S\ref{sec:Numerical-simulations} to compute
the streaming in the bulk through equation (\ref{eq:NS-bulk-streaming}).

\subsubsection*{Vertical walls:}

At $x=\left\{ 0,L_{x}\right\} $, the tangential velocity field is
\begin{equation}
\begin{array}{c}
\vs\uum_{0}=0,\\
\dps{(\uvm_{0},\uwm_{0})=\pm A\cos(\omega t)\left(-k_{2}\sin(k_{2}y)\frac{\cosh k(z+h)}{\cosh kh},k\cos(k_{2}y)\frac{\sinh k(z+h)}{\cosh kh}\right)},
\end{array}\label{eq-x=00003D0}
\end{equation}
which is used to get the matched streaming boundary conditions by
applying (\ref{eq-streaming_u1_vect2}) 
\begin{equation}
\begin{array}{c}
\vs\langle\uum_{1}\rangle=0,\\
\dps{(\langle\uvm_{1}\rangle,\langle\uwm_{1}\rangle)=\frac{A^{2}}{8\omega C_{0}^{2}}\left(k_{2}\sin(2k_{2}y)\left[3k_{1}^{2}C_{z}^{2}-k^{2}\right],-kS_{2z}\left[3k_{1}^{2}\cos^{2}(k_{2}y)+k_{2}^{2}\right]\right)}.
\end{array}\label{eq-x=00003D0_streamingBC}
\end{equation}
Here we have compacted the notations ${\cosh(k(z+h))}$, ${\sinh(k(z+h))}$
and ${\sinh(2k(z+h))}$ into $C_{z}$, $S_{z}$ and $S_{2z}$, respectively
($C_{0}$ stands for ${\cosh(kh)}$). By permuting the coordinates
$x\leftrightarrow y$, components $u\leftrightarrow v$ and indices
$_{1}\leftrightarrow\,_{2}$, we deduce the matched streaming boundary
conditions at the two other walls $y=\left\{ 0,L_{y}\right\} $. 
\begin{equation}
\begin{array}{c}
\vs\dps{(\langle\uum_{1}\rangle,\langle\uwm_{1}\rangle)=\frac{A^{2}}{8\omega C_{0}^{2}}\left(k_{1}\sin(2k_{1}x)\left[3k_{2}^{2}C_{z}^{2}-k^{2}\right],-kS_{2z}\left[3k_{2}^{2}\cos^{2}(k_{1}x)+k_{1}^{2}\right]\right)},\\
\langle\uvm_{1}\rangle=0.
\end{array}\label{eq-y=00003D0_streaming}
\end{equation}

\subsubsection*{Bottom:}

The bottom of the tank corresponds to $z=-h$. The instantaneous velocity
fields read there 
\begin{equation}
(\uum_{0},\uvm_{0})=-A\cos(\omega t)\left(k_{1}\frac{\sin(k_{1}x)\cos(k_{2}y)}{\cosh kh},k_{2}\frac{\cos(k_{1}x)\sin(k_{2}y)}{\cosh kh}\right)\label{eq-z=00003D-h}
\end{equation}
and the resulting matched boundary conditions are 
\begin{equation}
\begin{array}{c}
\vs\dps{(\langle\uum_{1}\rangle,\langle\uvm_{1}\rangle)=-\frac{A^{2}}{8\omega C_{0}^{2}}\left(k_{1}\sin(2k_{1}x)\left[3k^{2}\cos^{2}(k_{2}y)-k_{2}^{2}\right],k_{2}\sin(2k_{2}y)\left[3k^{2}\cos^{2}(k_{1}x)-k_{1}^{2}\right]\right)},\\
\langle\uwm_{1}\rangle=0.
\end{array}\label{eq-z=00003D-h_streamingBC}
\end{equation}

\subsubsection*{Top interface:}

At the interface ($z=0$), we have chosen two types of boundary conditions
that can be applied to the streaming field. The first one is 
\begin{equation}
\begin{array}{c}
\vs\ddz\langle\uum_{1}\rangle=\ddz\langle\uvm_{1}\rangle=0,\\
\langle\uwm_{1}\rangle=0,
\end{array}\label{eq-z=00003D0_streamingBC1}
\end{equation}
for uncontaminated surfaces. This condition is equivalent to the classic
free-surface condition, i.e. null tangential stress at $z=0$ \citep{Batchelor2000an_introduction}.
The second one is 
\begin{equation}
\begin{array}{c}
\vs\langle\uwm_{1}\rangle=0,\\
\dps{(\langle\uum_{1}\rangle,\langle\uvm_{1}\rangle)=-\frac{A^{2}}{8\omega}\left(k_{1}\sin(2k_{1}x)\left[3k^{2}\cos^{2}(k_{2}y)-k_{2}^{2}\right],k_{2}\sin(2k_{2}y)\left[3k^{2}\cos^{2}(k_{1}x)-k_{1}^{2}\right]\right)},
\end{array}\label{eq-z=00003D0_streamingBC2}
\end{equation}
for fully contaminated surfaces. This condition is due to the presence
of an inextensible film at the surface \citep{Miles:1967vp,1994JFM...275..285H,2005JFM...546..203M}.

\subsubsection{Case of longitudinal waves: $k_{2}=0$ \label{sub:longitudinal_waves}}

We now focus on the case of plane stationary waves characterised by
$k_{2}=0$, which corresponds to the striped Faraday-waves patterns
that can easily be observed in containers whose transverse dimension
is much smaller than the critical wavelength. Waves in this case are
almost two-dimensional. Nevertheless, some three-dimensional effects
may still be present due to the presence of the walls in the transverse
direction, which induces alteration of the streaming flow through
its viscosity. The expression for the instantaneous potential $\Phi$
is thus changed to 
\begin{equation}
\Phi(x,y,z,t)=A\cos(kx)\frac{\cosh k(z+h)}{\cosh kh}\cos(\omega t),\label{eq-potential_k2=00003D0}
\end{equation}
where naturally $k_{1}=k$. As a result, the matched boundary conditions
for the streaming velocities are also simplified. The results are
summarised in table~\ref{tab:BC-longitudinal}.

\begin{table}
\begin{centering}
\begin{tabular}[t]{rl>{\centering}p{2.5cm}>{\centering}p{2cm}>{\centering}p{2.5cm}}
 &
 &
$\vs\langle\uum_{1}\rangle$ &
$\vs\langle\uvm_{1}\rangle$ &
$\vs\langle\uwm_{1}\rangle$\tabularnewline
Bottom: &
$z=-h$ &
$-\frac{3}{8}\frac{A^{2}k^{3}}{\omega C_{0}^{2}}\sin(2kx)$ &
$0$ &
$0$\tabularnewline
Lateral walls: &
$x=\left\{ 0,L_{x}\right\} $ &
$0$ &
$0$ &
$-\frac{3}{8}\frac{A^{2}k^{3}}{\omega C_{0}^{2}}\sinh\left(2k\left(z+h\right)\right)$\tabularnewline
Front walls: &
$y=\left\{ 0,L_{y}\right\} $ &
$-\frac{1}{8}\frac{A^{2}k^{3}}{\omega C_{0}^{2}}\sin(2kx)$ &
$0$ &
$-\frac{1}{8}\frac{A^{2}k^{3}}{\omega C_{0}^{2}}\sinh\left(2k\left(z+h\right)\right)$\tabularnewline
Top interface: &
$z=0$$^{*}$ &
$-\frac{3}{8}\frac{A^{2}k^{3}}{\omega}\sin(2kx)$ &
$0$ &
$0$\tabularnewline
 &
$z=0$$^{\dagger}$ &
$\ddz\vs\langle\uum_{1}\rangle=0$ &
$\ddz\vs\langle\uvm_{1}\rangle=0$ &
$0$\tabularnewline
\end{tabular}
\par\end{centering}

\protect\caption{Streaming matched boundary conditions for longitudinal Faraday waves
($k=k_{1},\, k_{2}=0$). Only the conditions at the top interface
change for a fully contaminated interface ($^{*}$) and an uncontaminated
one ($^{\dagger}$). These boundary conditions are used in the numerical
simulations presented in \S\ref{sec:Numerical-simulations}. \label{tab:BC-longitudinal}}
\end{table}

\subsubsection{Two-dimensional flow\label{sub:Two-dimensional-flow}}

When the problem is completely homogeneous in the $y$ direction (absence
of transverse walls, $\ddy$=0, $v=\langle\uvm_{1}\rangle=0$), the
potential takes the same expression as in the case $k_{2}=0$ (\ref{eq-potential_k2=00003D0}).
The matched boundary conditions in table~\ref{tab:BC-longitudinal}
remain similar while those at the transverse walls drop. The transverse
walls are the only source of differences between the $k_{2}=0$ and
the two-dimensional flow. In the former case, the presence of the
wall is expected to induce a three-dimensional streaming field.

\section{Numerical simulations\label{sec:Numerical-simulations}}

In order to solve equations (\ref{eq:NS-bulk-streaming}) with the
matched boundary conditions from table~\ref{tab:BC-longitudinal}
provided by the theory, we require numerical simulations. They allow
us to address two basic questions: Can we reproduce experimental observations
using the boundary-layer-induced streaming theory? Is it sufficient
to keep a two-dimensional approximation or do three-dimensional effects
play an important role?

\subsection{Numerical methods}

To run simulations of the streaming flow inside the incompressible
vibrating fluid, we implemented our own code. The code directly integrates
the Navier-Stokes equations with the divergence-free condition (\ref{eq:NS-bulk-streaming})
in a parallelepiped domain delimited at the top by the interface of
the fluid at rest (see figure~\ref{fig:bords}(\emph{b})). The flat
top interface is a fair approximation for the time-averaged position
of low amplitude Faraday waves. In (\ref{eq-NS2_in}), the gravity
and the vertical vibration are gathered with the pressure to form
a time-dependent effective pressure and do not have any effect on
the velocity fields since a single fluid of constant density is involved
in this model. In the code, the fields are discretised by finite-difference
schemes using the MAC (Marker And Cell) disposition on a regular staggered
mesh where the pressure is computed at the centre of each cell and
the components of the velocity are located at the faces corresponding
to their direction. The incompressible Navier-Stokes equations are
solved by a projection method \citep{Chorin:1968uj}. The pressure
is computed incrementally with a BiCGStab (Stabilised BiConjugate
Gradient) method. The spatial discretisation is made with standard
centred schemes of second order except the advection terms for which
ENO (Essentially Non Oscillatory) schemes are used. The temporal stepping
involves a first-order forward Euler discretisation with adaptive
time steps to ensure numerical stability.

The simulations use the equations from table~\ref{tab:BC-longitudinal}
as matched boundary conditions, with the three quantities, $k$, $\omega$
and $A$, fixed. We recall that for longitudinal waves the wave number
is defined as $k=m\upi/L_{x}$ with $m$ an integer, and $\omega$
is deduced from $k$ by means of (\ref{eq:dispersion}) in consistency
with the experiments. At dominant order, the amplitude $A$ is related
with the amplitude of Faraday waves $\Delta\zeta=\max_{x,t}(\zeta)-\min_{x,t}(\zeta)$
by 

\begin{equation}
A\sim\frac{g\Delta\zeta}{\omega}.\label{eq-Astream_Amplitude-1}
\end{equation}

We emphasize some features about the theoretical model and the simulations.
First, the fact that the coefficients $k$, $\omega$ and $A$ are
set up manually renders the model and the simulations independent
of the bifurcation criticality, a feature consistent with the experimental
results. Second, all the matched boundary conditions for the streaming
flow are proportional to $A^{2}/\omega$. Hence, cases with altered
$\omega$ are also swept by varying $A$; the complete model could
be rewritten as a function of the single variable $A'=A/\sqrt{\omega}$
instead of $\omega$ and $A$ separately, reducing the dimension of
the parameter space.

\subsection{Numerical results\label{sub:Numerical-results}}

\begin{figure}
\begin{centering}
\includegraphics[width=1\textwidth]{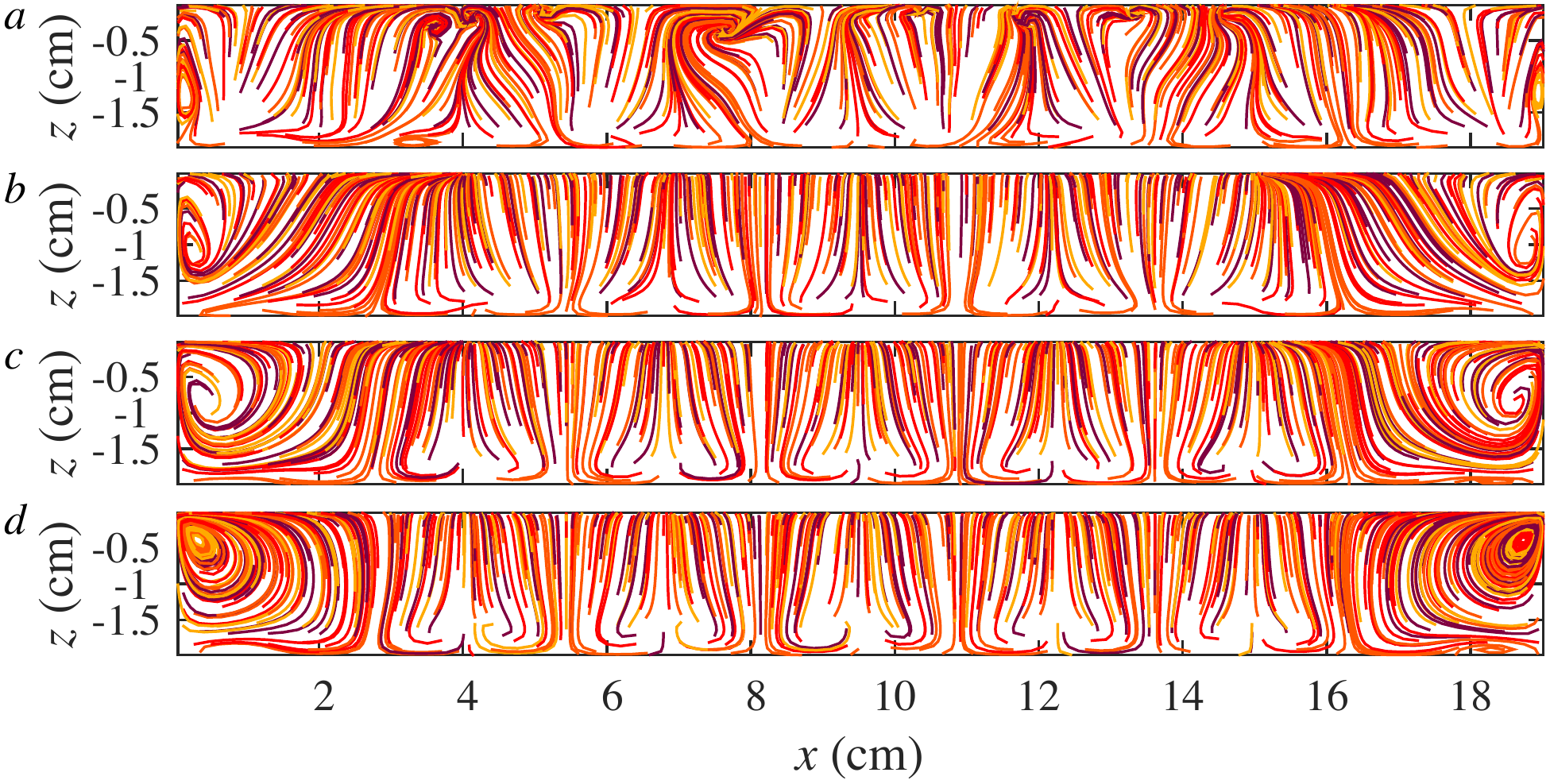}
\par\end{centering}

\protect\caption{Instantaneous streamlines in the $y$ midplane for $m=7$ and varying
$A$ with uncontaminated interface. The amplitude $A$ increases from
the bottom to the top: (\emph{a}) $71.1$, (\emph{b}) $40$, (\emph{c})
$22.5$ and (\emph{d}) $0.4\,\mathrm{cm^{2}/s}$.}

\label{fig:LC-NC}
\end{figure}

\begin{figure}
\begin{centering}
\includegraphics[width=1\textwidth]{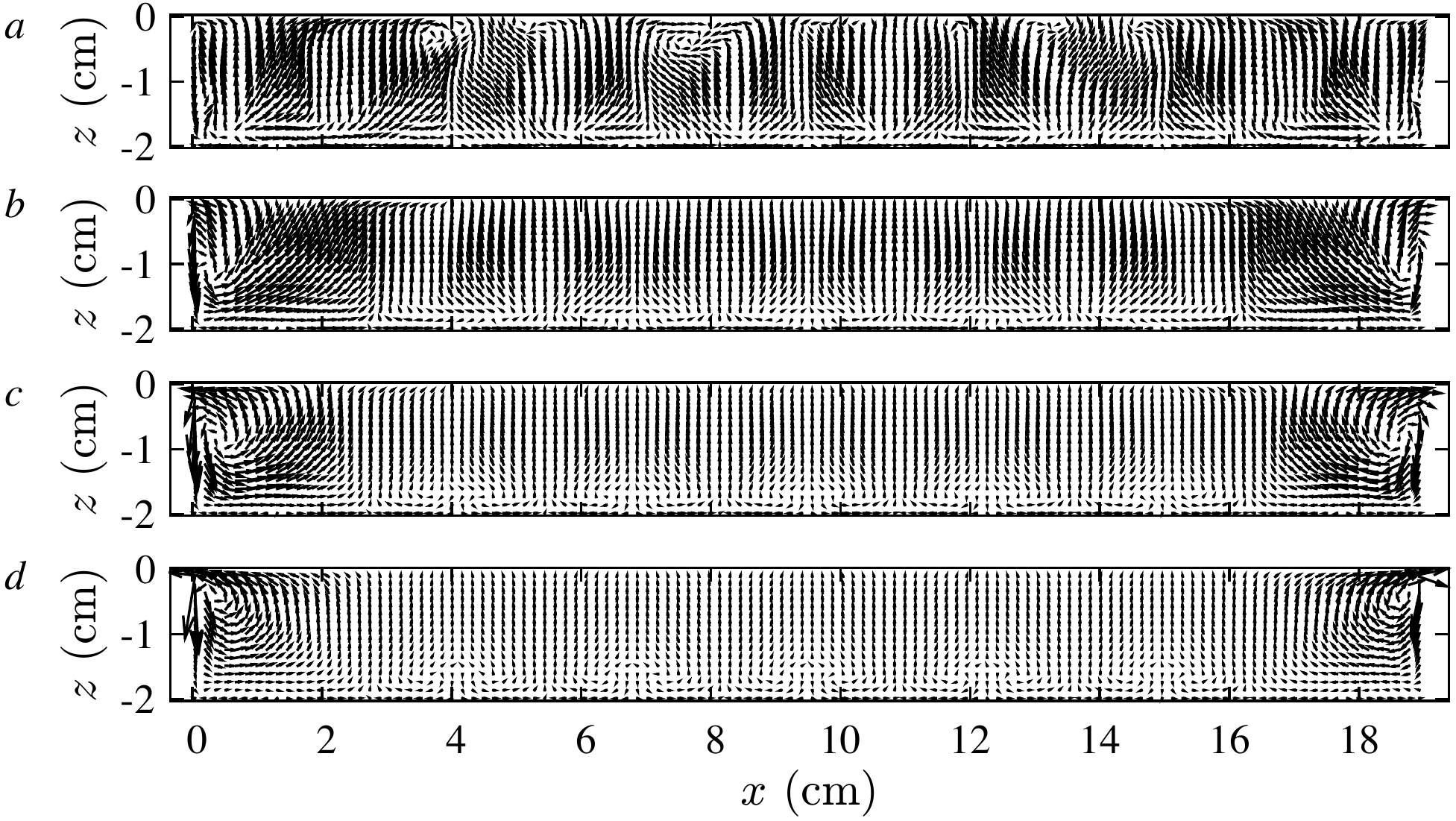}
\par\end{centering}

\centering{}\protect\caption{Instantaneous velocity field in the $y$ midplane for $m=7$ and varying
$A$ with uncontaminated interface. Amplitudes are the same as in
figure \ref{fig:LC-NC}.}
\label{fig:vel_field_uncont}
\end{figure}

In this section, most of the parameters of interest are the same as
in the experiment: the density $\rho=1.1$ g/cm$^{3}$, the dynamic
viscosity $\nu=9.09\times10^{-3}\,\mathrm{cm^{2}/s}$, the dimensions
of the fluid layer $L_{x}\times L_{y}\times h=(19.05\times2.54\times2)$
cm$^{3}$ and the number of nodes in the longitudinal direction $m=4,5,6,7$.
The corresponding frequencies of Faraday waves given by the dispersion
relation (\ref{eq:dispersion}) are $f/2=$ 3.8 Hz, 4.45 Hz, 4.95
Hz and 5.4 Hz, respectively (the frequency of vibration of the container
corresponds to $f$ as shown in table~\ref{tab:Measurements}). The
amplitude $A$ is varied between $0.4\,\mathrm{cm^{2}/s}$ and $1.26\times10^{2}\,\mathrm{cm^{2}/s}$
, where the lowest bound is the value below which the streaming flow
remains qualitatively unchanged and the upper bound is the value at
which the code diverges. The regular discretised mesh contains $240\times80\times80$
cells. Simulations are performed with both the uncontaminated and
the fully contaminated boundary conditions at the interface, cf. table~\ref{tab:BC-longitudinal}.
In figures~\ref{fig:LC-NC}--\ref{fig:vel_field_uncont}, we first
show the instantaneous streamlines and associated velocity field of
the streaming flow in the $y$ midplane. We set $m=7$, $f/2=5.4$
Hz on an uncontaminated interface as $A$ is varied.

\begin{figure}
\begin{centering}
\includegraphics[width=1\textwidth]{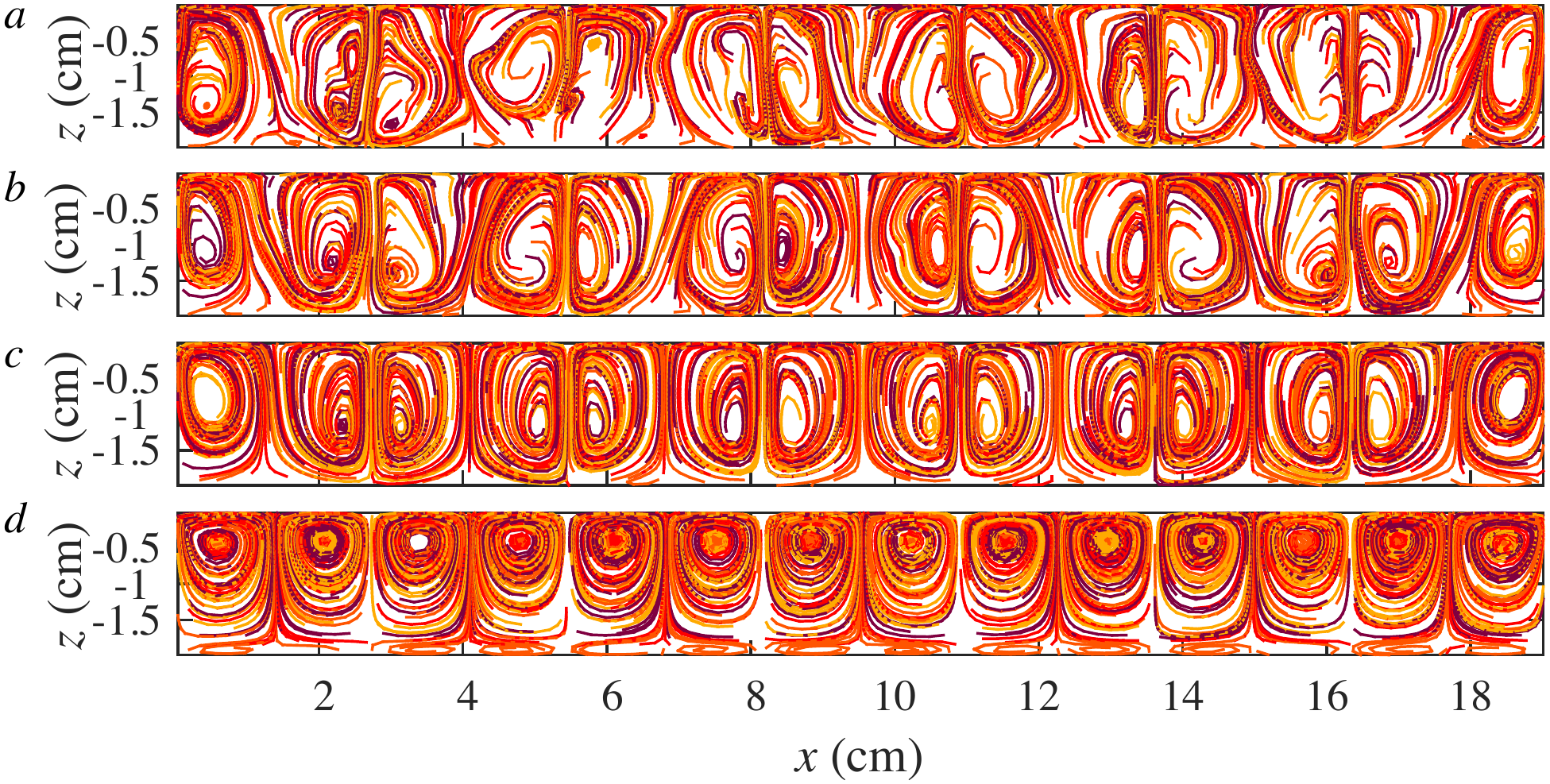}
\par\end{centering}

\protect\caption{Instantaneous streamlines in the $y$ midplane for $m=7$ and varying
$A$ with fully contaminated interface. Same amplitudes as in figure
\ref{fig:LC-NC}.}

\label{fig:LC-C}
\end{figure}

\begin{figure}
\begin{centering}
\includegraphics[width=1\textwidth]{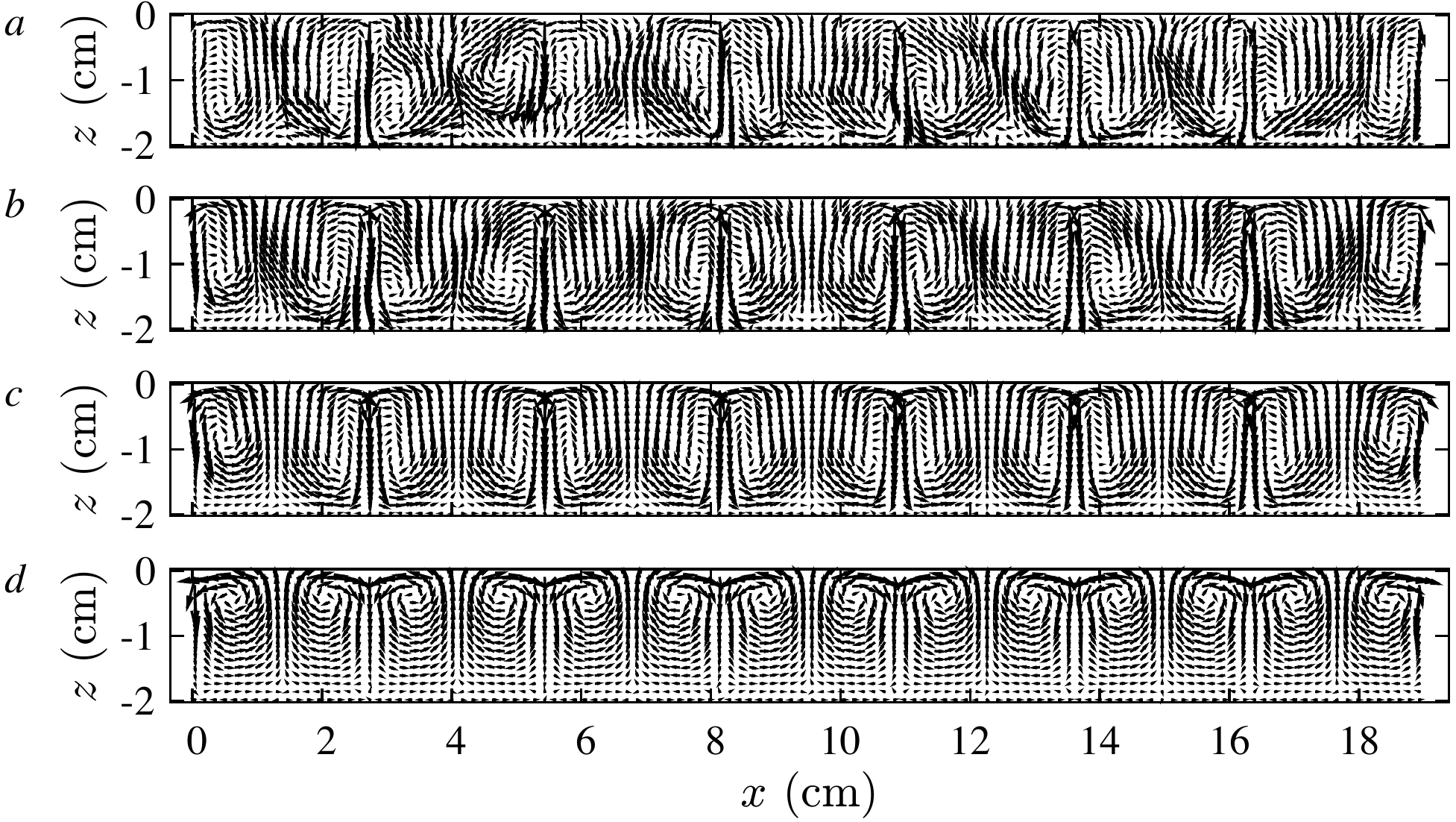}
\par\end{centering}

\protect\caption{Instantaneous velocity field in the $y$ midplane for $m=7$ and varying
$A$ with fully contaminated interface. Same amplitudes as in figure
\ref{fig:LC-NC}.}

\label{fig:vitesses-C-plan_y}
\end{figure}

Likewise, in figures \ref{fig:LC-C}--\ref{fig:vitesses-C-plan_y},
we plot respectively the instantaneous streamlines and the associated
velocity fields of the Faraday streaming flow for the same values
of $m$ and $\omega$ but for the fully contaminated case. 

Comparison of figures \ref{fig:LC-NC}--\ref{fig:vel_field_uncont}
with \ref{fig:LC-C}--\ref{fig:vitesses-C-plan_y} show that contamination
induces very unlike streaming flows at low amplitudes. The low-amplitude
fully contaminated Faraday streaming flows are very similar to the
type-I patterns found in the experiments, with weakly spiraling loops
at the top of the domain and small recirculations at the bottom. The
fact that the streaming flows do not form perfectly closed loops is
related to the presence of transverse variations of the velocity component
$v$. The features of the flow are then expected to differ from those
obtained with the two-dimensional theory, even on the $y$ midplane.
The experimental patterns of type I (figures \ref{fig:overexposed}--\ref{fig:piv},
patterns (\emph{c}) and (\emph{d})) display the strongest similarities
with the simulated fields of figures \ref{fig:LC-C}--\ref{fig:vitesses-C-plan_y}
at moderate amplitude $A=22.5\,\mathrm{cm^{2}/s}$. Therein, both
the streamlines and the velocity fields have very similar shapes.

The uncontaminated Faraday streaming flows look completely different
from the fully contaminated ones and contrast with all the patterns
observed in the experiments. However, they display similarities with
the type-II patterns, namely the presence of accumulation points at
the vicinity of the interface, suggesting again that the flow is fairly
three-dimensional on the $y$ midplane. In contrast, there are no
loops at the top of the domain. Intermediate contamination, or in
other words matched boundary conditions between (\ref{eq-z=00003D0_streamingBC1})
and (\ref{eq-z=00003D0_streamingBC2}) as in (\citealt{2002JFM...467...57M}),
might improve the resemblance between experiments and simulations.

Examples of disordered Faraday streaming flows are shown at larger
amplitudes like $A=71.1\,\mathrm{cm^{2}/s}$ at the top of each figure
\ref{fig:LC-NC}--\ref{fig:vitesses-C-plan_y}. When disorder comes,
streaming flows become unsteady as well. The emergence of disorder
generally occurs at values of $A$ of that order of magnitude, regardless
the interface contamination and for all the wave numbers explored
in this work.

Varying the wave number $m$ in the matched boundary conditions does
not have any dramatic effect in the flow structure. Besides the expected
change of wavelength of the streaming flows that occurs, one can notice
the loops at the walls are overcome by the central patterns as $m$
is decreased, particularly in the uncontaminated case.

\begin{figure}
\begin{centering}
\includegraphics[width=4in]{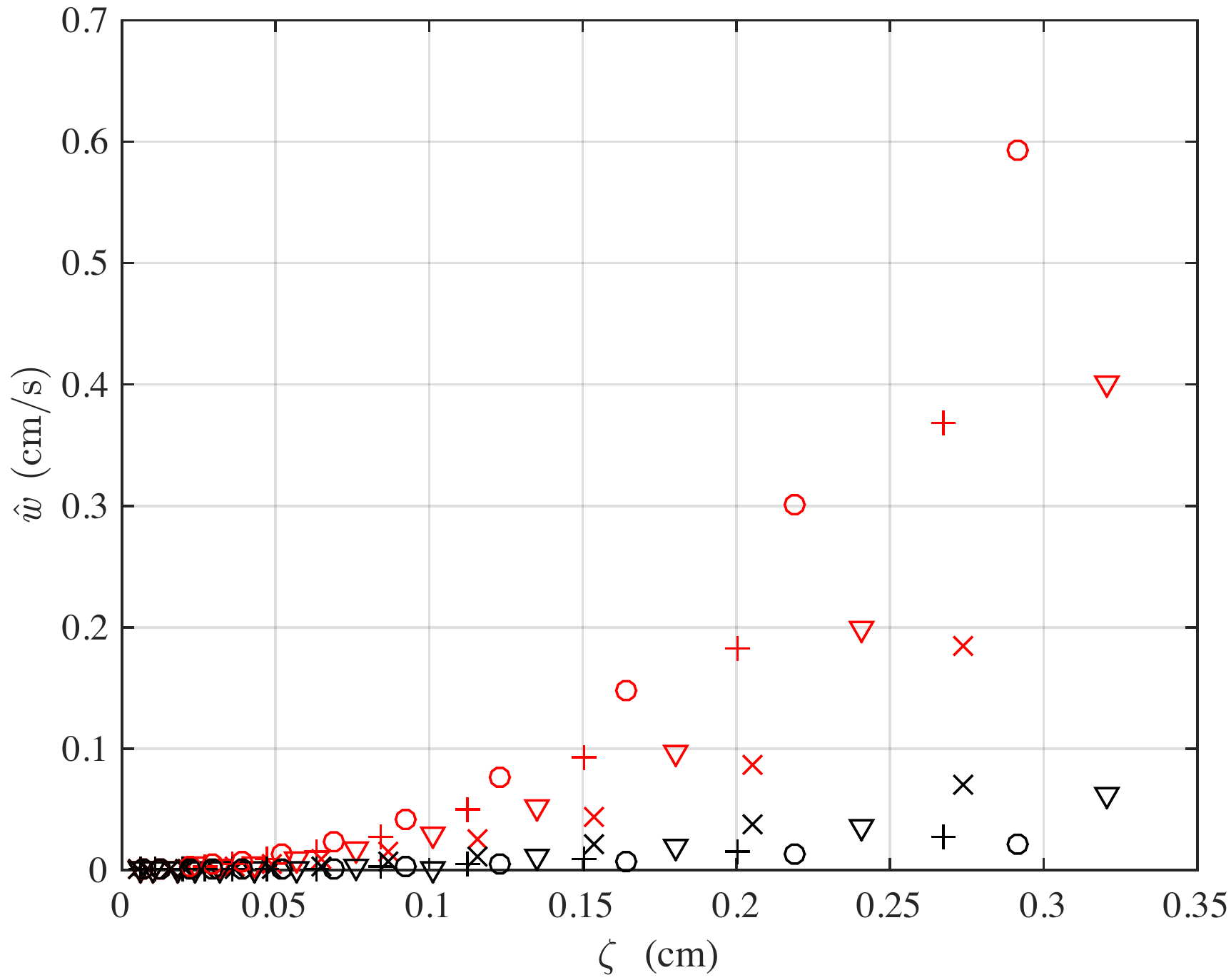}
\par\end{centering}

\protect\caption{Bifurcation diagram for simulations of streaming patterns. Red signs
represent contaminated Faraday streaming flows, which share most properties
of the Type-I patterns. Black signs stand for uncontaminated Faraday
streaming flows that are roughly similar to type-II patterns. The
markers $\times$, $\triangledown$, $+$ and $\ocircle$ correspond
to $m=$4, 5, 6 and 7, respectively.}

\label{fig:bifurcation_diagram_num-1}
\end{figure}

We then focus on more quantitative aspects of the Faraday streaming
flows to compare our simulations with the experimental results presented
in figure~\ref{fig:Bifurcation-diagram}. Taking the same definition
of the average velocity $\hat{w}$ as in $\S$\ref{sub:Experimental-results},
we plot in figure~\ref{fig:bifurcation_diagram_num-1} a comparable
bifurcation diagram. Different symbols correspond to different $m$
values (see caption), while different colours distinguish between
patterns obtained with contaminated surface (in red) from those obtained
with uncontaminated surface (in black). We recall that uncontaminated
patterns share some properties with type-II patterns, while contaminated
ones resemble type-I patterns. 

Figures~\ref{fig:Bifurcation-diagram} and \ref{fig:bifurcation_diagram_num-1}
show remarkable similarities. First, $\hat{w}$ increases faster than
linearly in $\zeta$. In the two cases, $\hat{w}$ displays small
values for $\zeta<0.15\,\mathrm{cm}$. We also observe fair agreement
as the uncontaminated Faraday streaming flows generate lower vertically
averaged velocities than contaminated Faraday streaming flows. Finally
The values of $\hat{w}$ obtained in simulations are remarkably close
to those measured in the experiments. However, there is an important
discrepancy between the numerical simulation and the experiments:
simulations do not show a collapse of the branches as the experiments,
but instead, scattered branches with an important dependence on $m$.
The origin of this disagreement is still an open issue. It may come
from unsufficiently controled surface properties in the experiment,
or from the absense of noise in simulations. 

\begin{figure}
\begin{centering}
\includegraphics[width=1\textwidth]{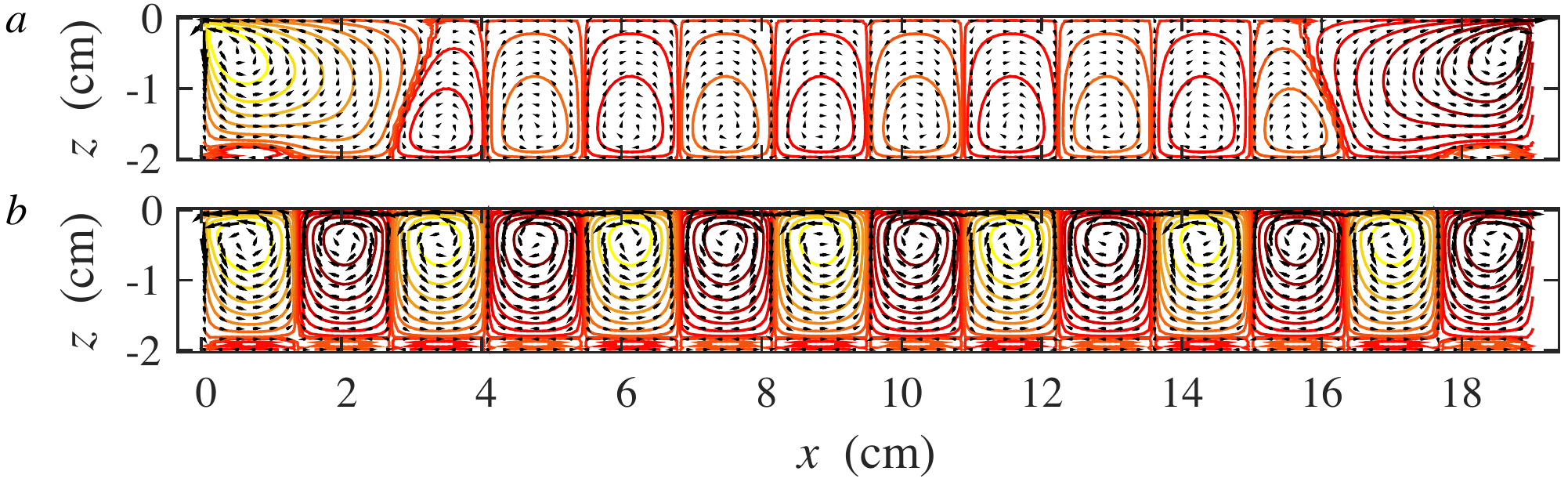}
\par\end{centering}

\protect\caption{Velocity fields (arrows) and streamlines (contours) of the two-dimensional
streaming flow for $m=7$ and $A=0.4\,\mathrm{cm^{2}/s}$. From top
to bottom: (\textit{a}) with uncontaminated interface and (\textit{b})
with fully contaminated interface.}

\label{fig:LC2D}
\end{figure}

To see whether the two-dimensional approximation used in previous
works is good enough to explain qualitatively the streaming field,
we exploit the tridimensional data obtained from the numerical simulations.
First, we compare the three-dimensional streaming patterns of our
numerical simulations in the $y$ midplane (figures \ref{fig:LC-NC}--\ref{fig:vitesses-C-plan_y})
to their two-dimensional counterparts (figure \ref{fig:LC2D}) at
low amplitude $A=0.4\,\mathrm{cm^{2}/s}$ and for $m=7$. For this
purpose, we ran a two-dimensional simplified version of the code following
\S\ref{sub:Two-dimensional-flow}. The main qualitative differences
between the streaming fields of figures \ref{fig:LC-NC}--\ref{fig:vitesses-C-plan_y}
and the two-dimensional ones of figure \ref{fig:LC2D} stem from the
fact that streamlines become perfectly closed loops when the flow
is purely two-dimensional, as expected. The differences are localized
far from the walls of the domain and are especially striking in the
uncontaminated case (compare figure \ref{fig:LC2D} (\textit{a}) with
figures \ref{fig:LC-NC} (\textit{d}) and \ref{fig:vel_field_uncont}
(\textit{d})). The differences remain visible when the interface is
contaminated at the frontier between two superposed loops, but in
a lesser extent (compare figure \ref{fig:LC2D} (\textit{b}) with
figures \ref{fig:LC-C} (\textit{d}) and \ref{fig:vitesses-C-plan_y}
(\textit{d})). Nevertheless, the main differences between the two
dimensional streaming flows and the three-dimensional ones are located
in regions where the streaming velocity is weak. 

\begin{figure}
\begin{centering}
\includegraphics[width=1\textwidth]{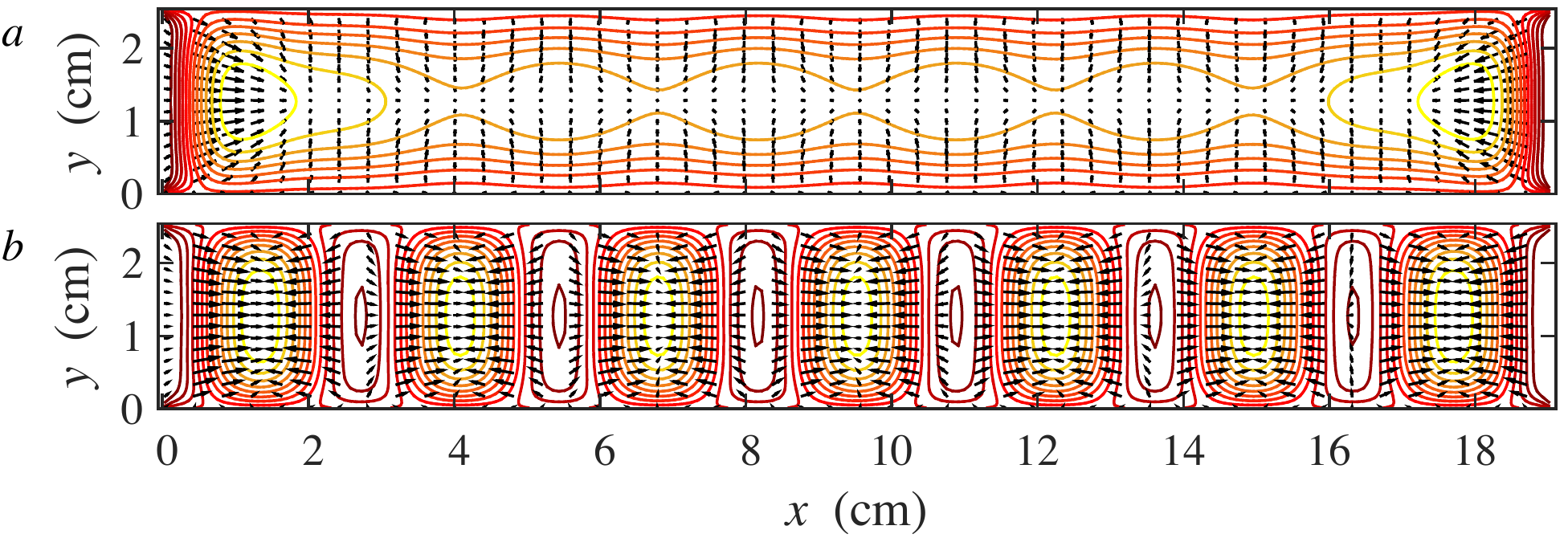}
\par\end{centering}

\protect\caption{Velocities of the streaming flow in the horizontal plane $z=-1$cm
for $m=7$ and $A=0.4\,\mathrm{cm^{2}/s}$. Arrows represent the horizontal
velocity components ($u,v$) while contours stand for $w$. From top
to bottom: (\textit{a}) with uncontaminated interface and (\textit{b})
with fully contaminated interface.}

\label{fig:v_z=00003D-1}
\end{figure}

Transverse effects are more visible in horizontal planes, for instance
at $z=-1\,\mathrm{cm}$ as shown in figure \ref{fig:v_z=00003D-1}.
They are highlighted by significant gradients of $w$ in the transverse
direction. Likewise, perceptible deviations of the velocity from or
towards the midplane $y=L_{y}/2$ are observed. The transverse effects
are more marked near the walls and exacerbate with increasing wave
amplitude $A$ and decreasing trough width $L_{y}$. These assessments
indicate that three-dimensional effects on Faraday streaming flows
are noticeable, even for longitudinal waves. This contrasts with the
weak and local effect of walls on Faraday surface waves.

\begin{figure}
\begin{centering}
\includegraphics[width=0.46\textwidth]{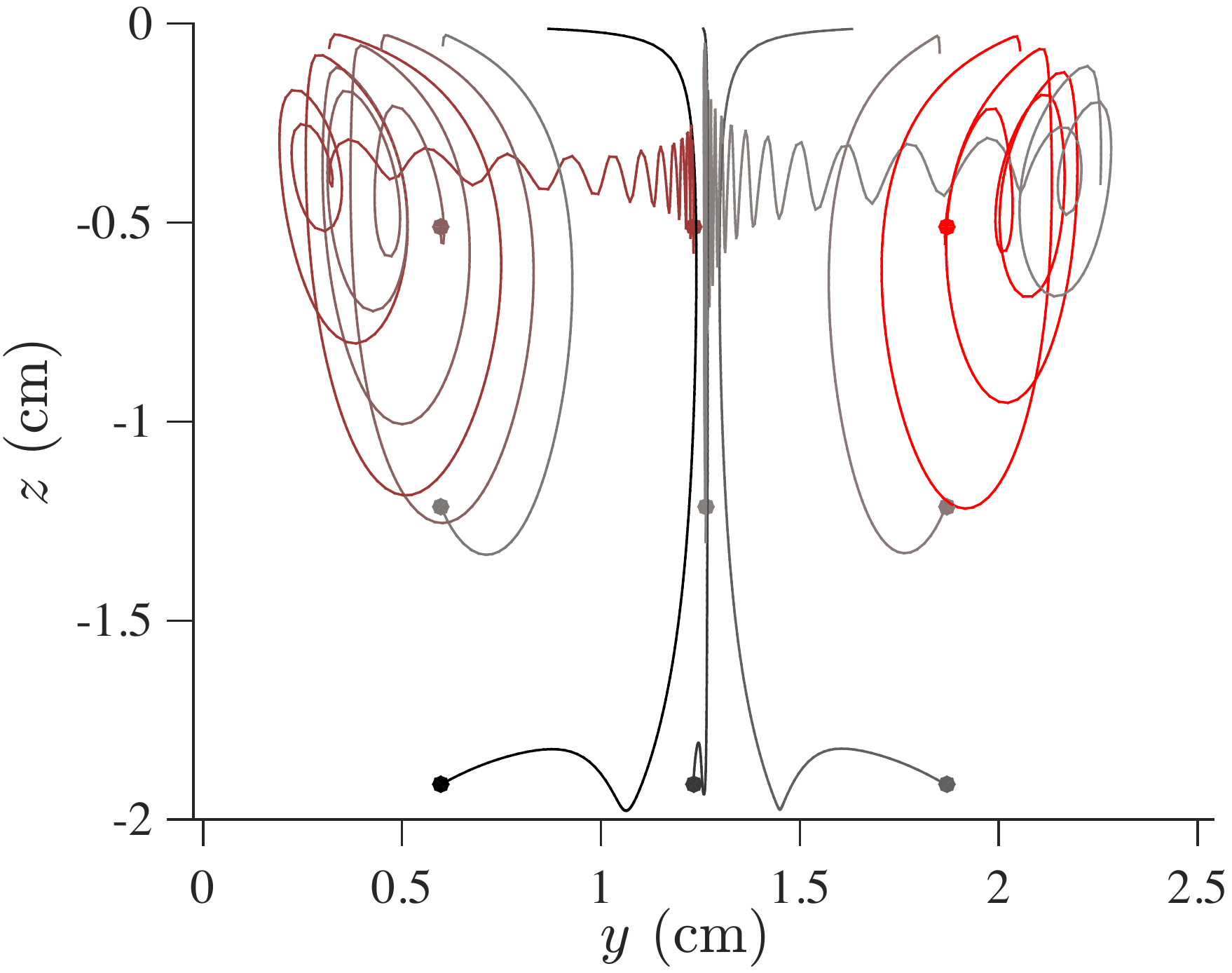}\includegraphics[width=0.52\textwidth]{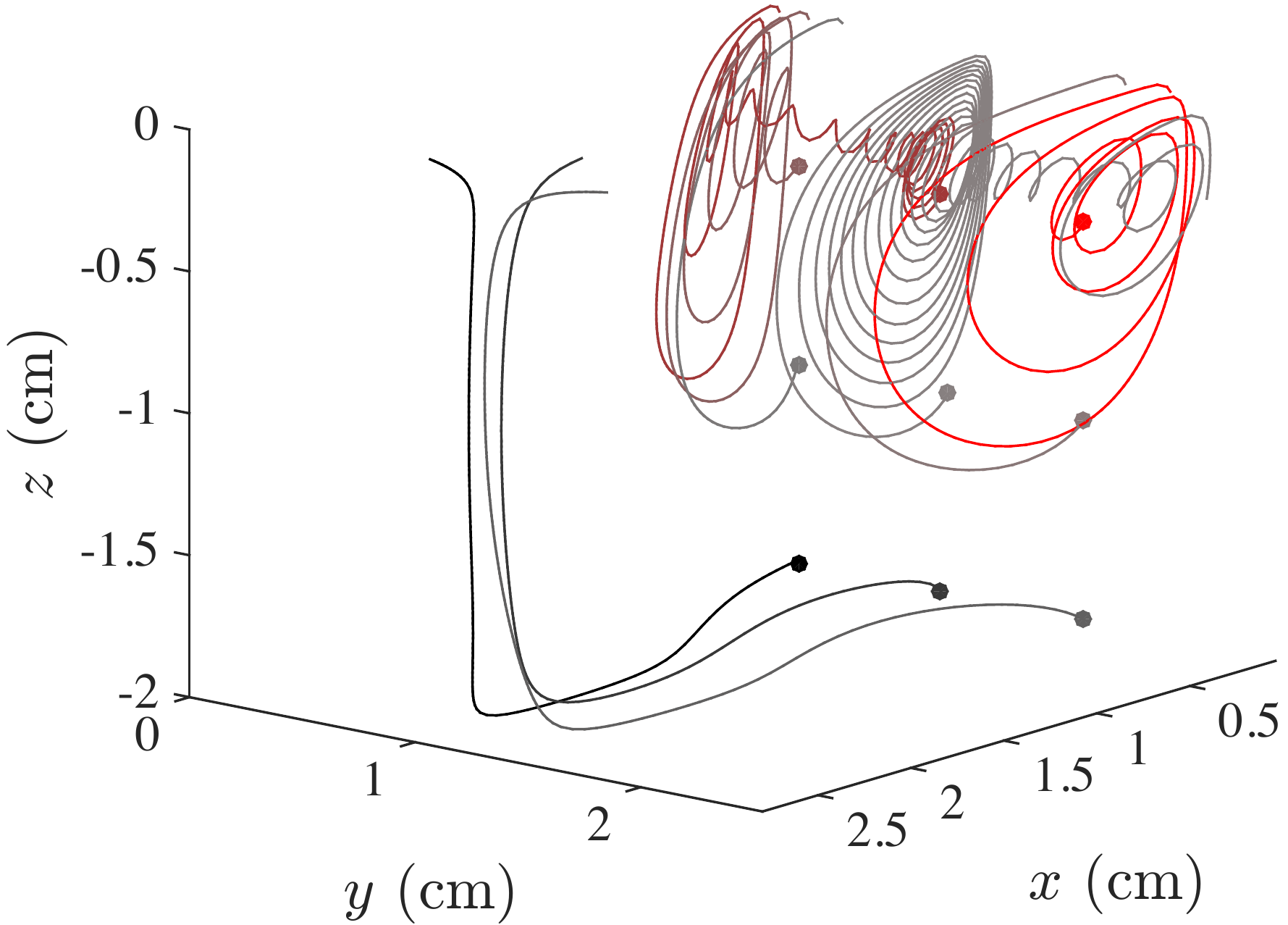}
\par\end{centering}

\protect\caption{Trajectories of particles starting from various points of the plane
$x=0.28$ cm (first cell) for $m=7$, $f/2=5.4$\foreignlanguage{english}{
Hz} and $A=0.4\,\mathrm{cm^{2}/s}$ with uncontaminated interface.
Left: viewed from the face $x=0$. Right: slanted view. Each color
denotes the trajectory of the same particle in left and right plots.
The starting points of the paths are represented by disks. \label{fig:trajectoires-NC-extremites}}
\end{figure}

\begin{figure}
\begin{centering}
\includegraphics[width=0.54\textwidth]{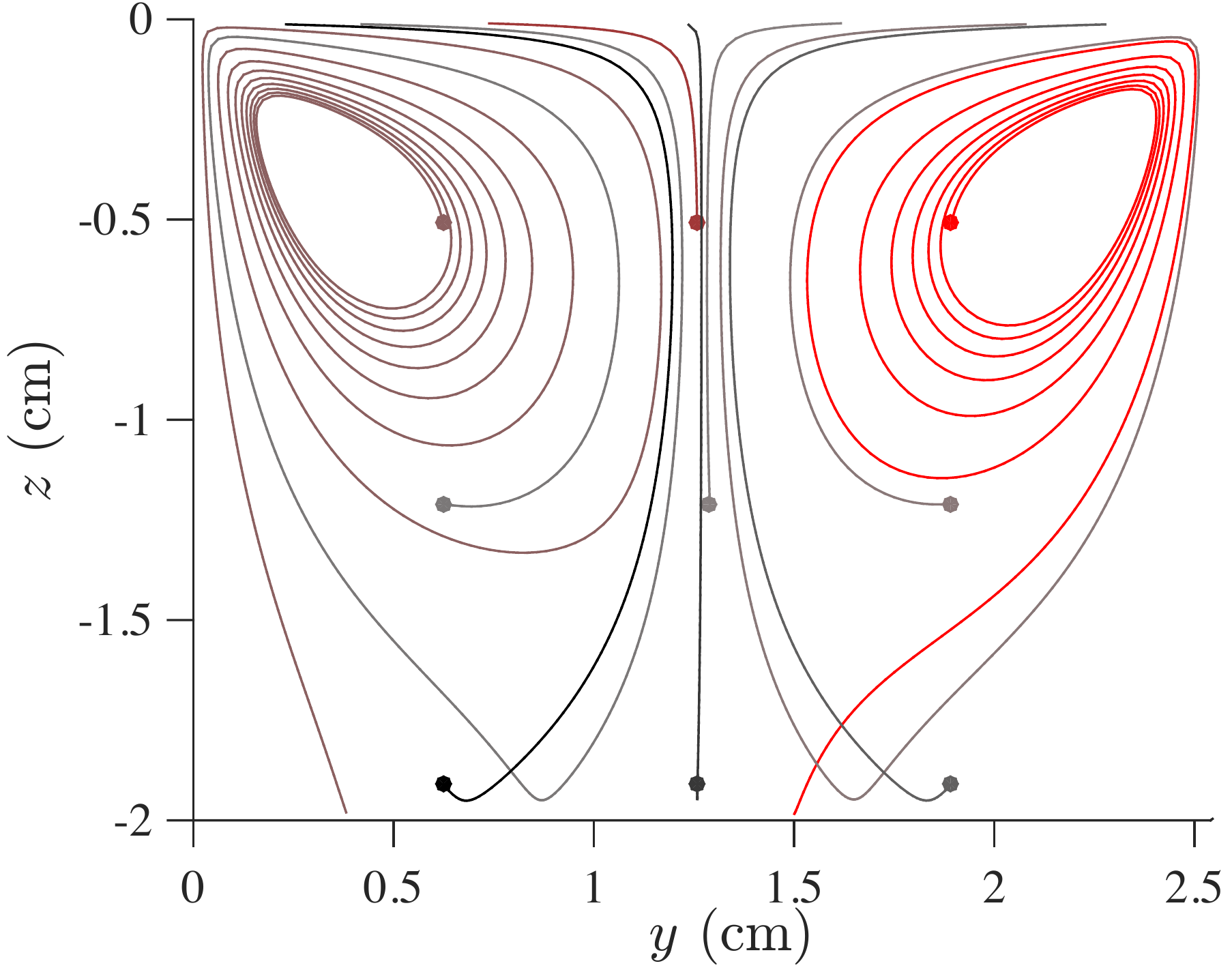}\includegraphics[width=0.42\textwidth]{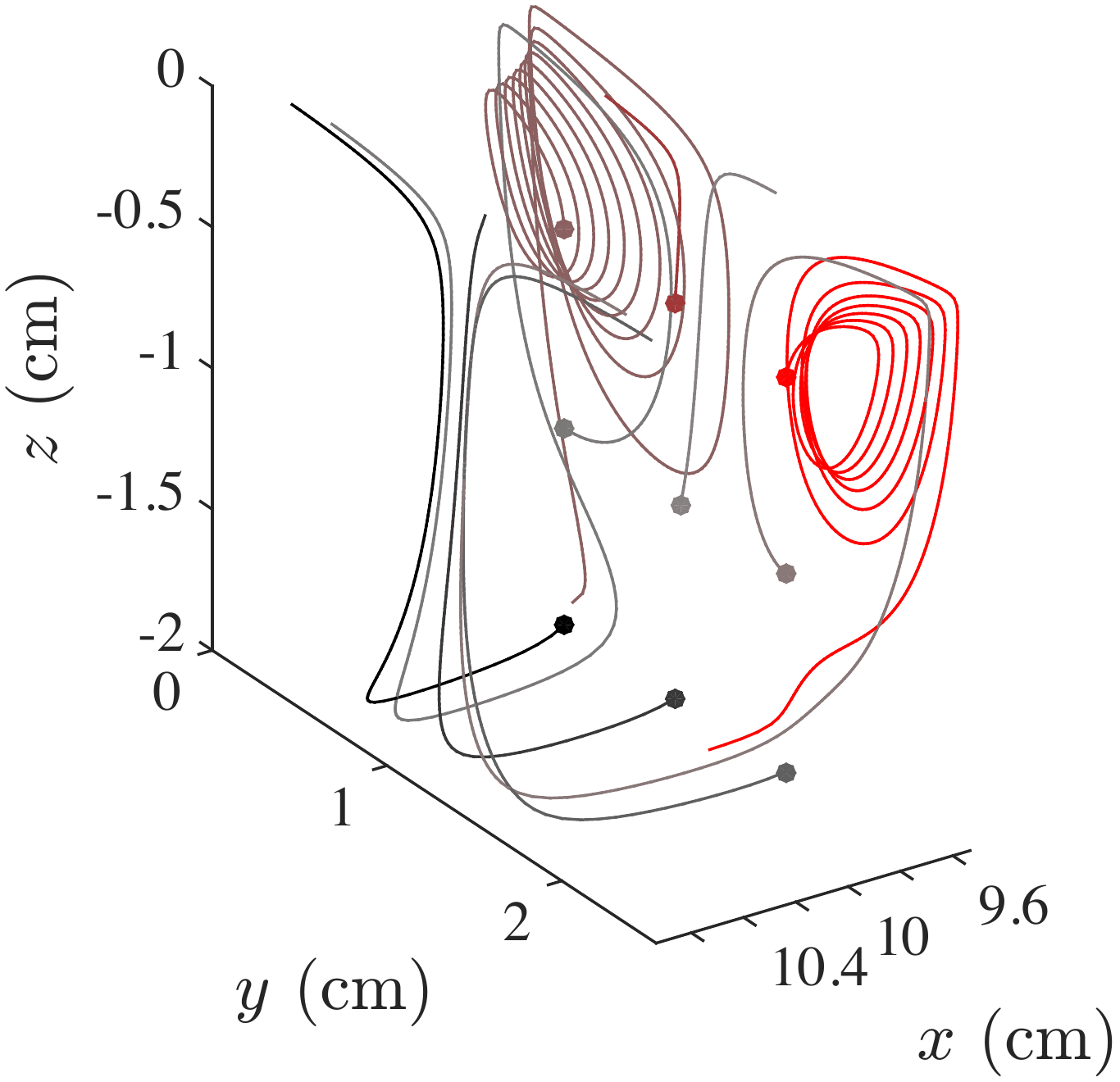}
\par\end{centering}

\protect\caption{Trajectories of particles starting from various points of the plane
$x=9.80$ cm (central cell) for $m=7$, $f/2=5.4$ Hz and $A=0.4\,\mathrm{cm^{2}/s}$
with uncontaminated interface. Left: viewed from the face $x=0$.
Right: slanted view. Each color denotes the trajectory of the same
particle in left and right plots. The starting points of the paths
are represented by disks. \label{fig:trajectoires-NC-centre}}
\end{figure}

\begin{figure}
\begin{centering}
\begin{minipage}[t]{0.49\columnwidth}%
\begin{center}
\includegraphics[width=1\textwidth]{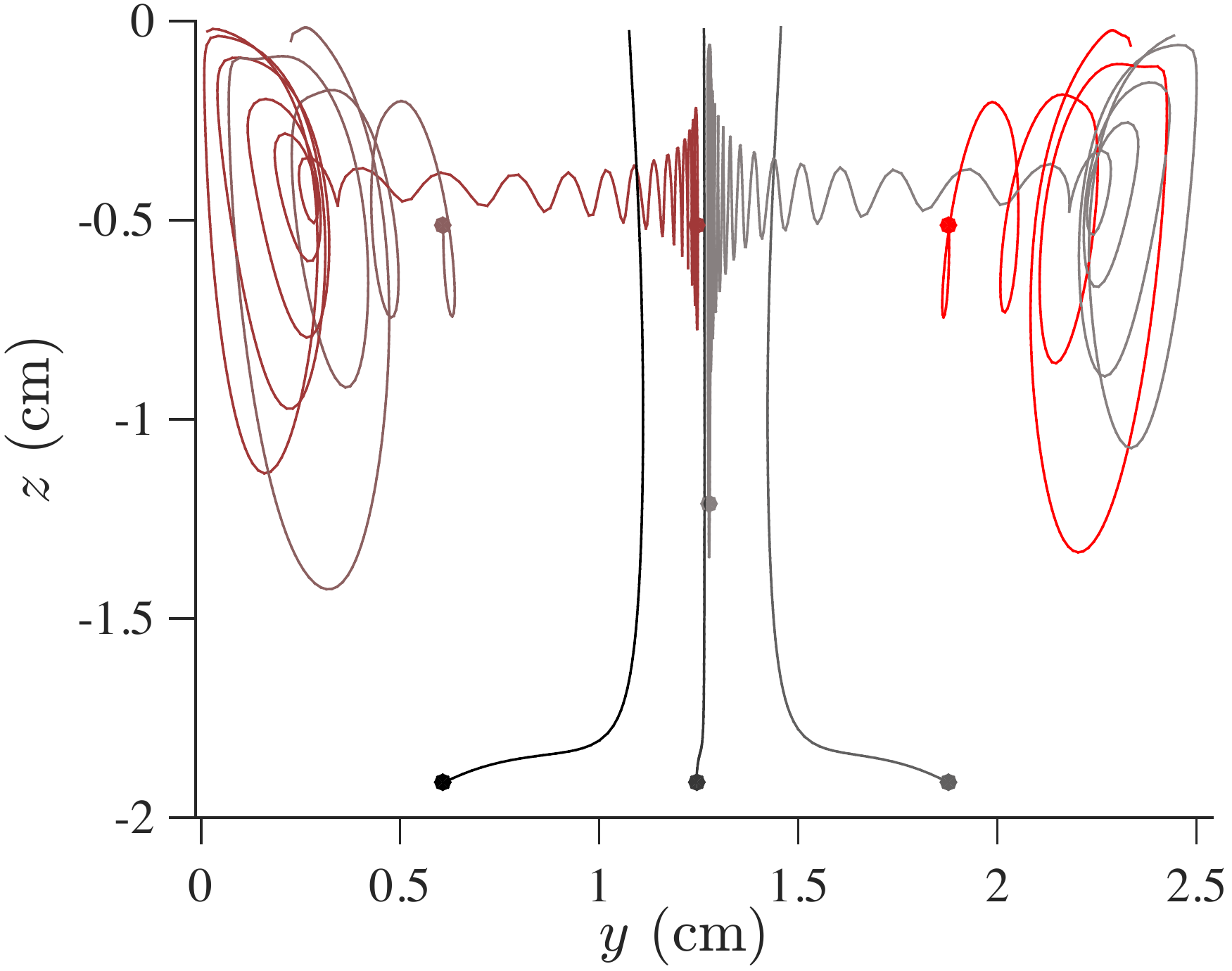}
\par\end{center}%
\end{minipage}%
\begin{minipage}[t]{0.49\columnwidth}%
\begin{center}
\includegraphics[width=1\textwidth]{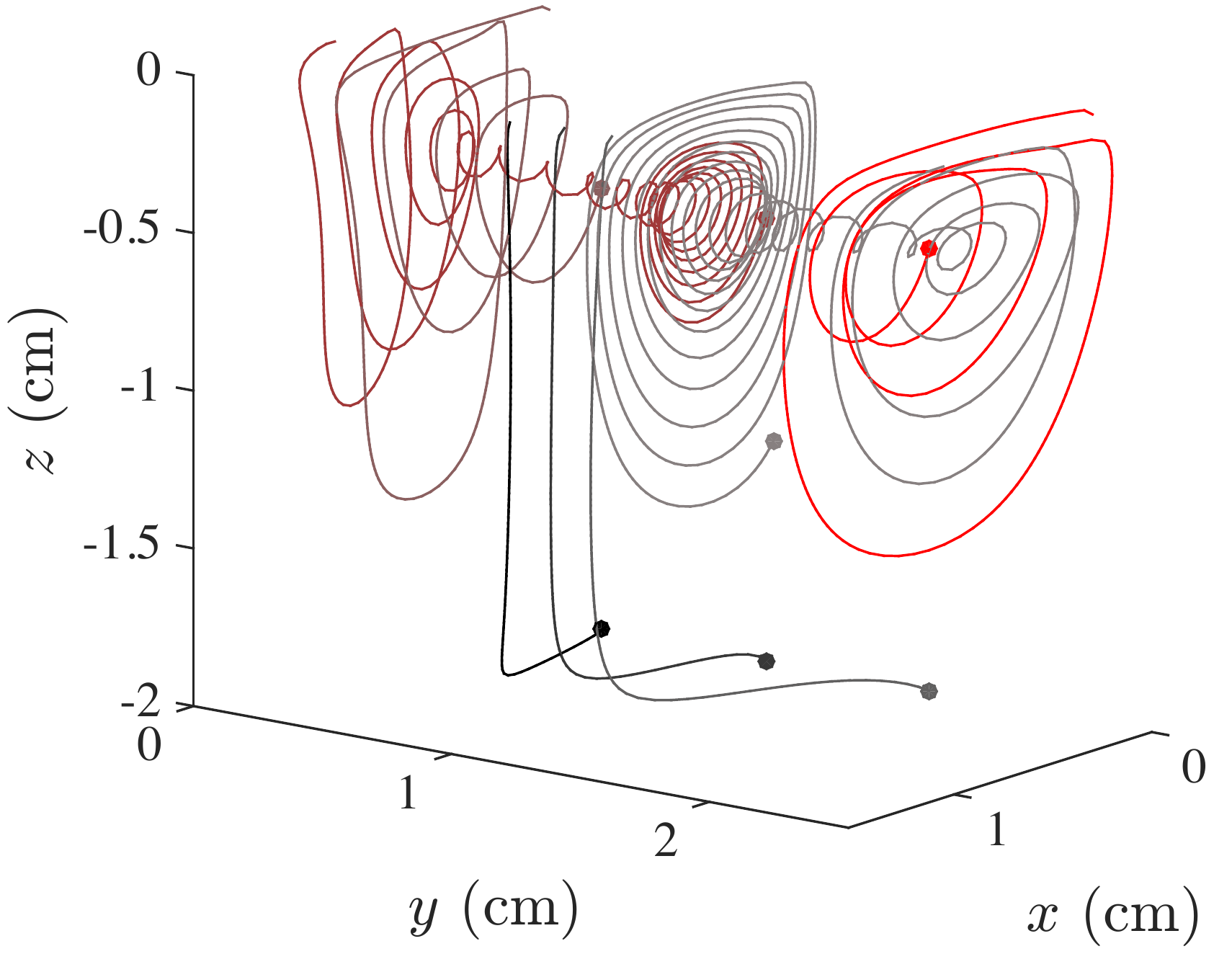}
\par\end{center}%
\end{minipage}
\par\end{centering}

\protect\caption{Trajectories of particles starting from various points of the plane
$x=0.28$ cm (first cell) for $m=7$, $f/2=5.4$\foreignlanguage{english}{
Hz} and $A=0.4\,\mathrm{cm^{2}/s}$ with fully contaminated interface.
Left: viewed from the face $x=0$. Right: slanted view. Each color
denotes the trajectory of the same particle in left and right plots.
The starting points of the paths are represented by disks.\label{fig:trajectoires-C-extremites}}
\end{figure}

\begin{figure}
\begin{centering}
\begin{minipage}[t]{0.54\columnwidth}%
\begin{center}
\includegraphics[width=1\textwidth]{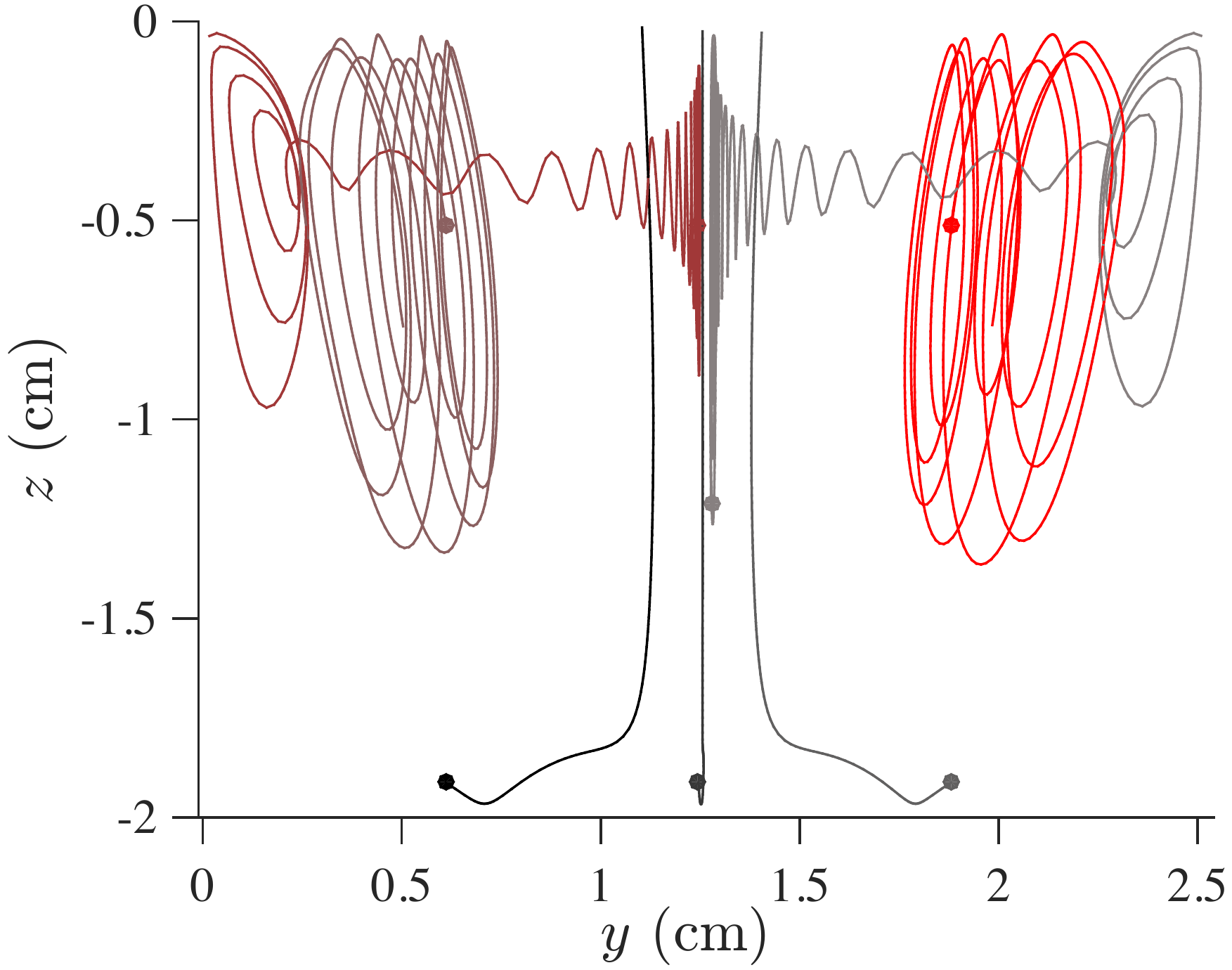}
\par\end{center}%
\end{minipage}%
\begin{minipage}[t]{0.44\columnwidth}%
\begin{center}
\includegraphics[width=1\textwidth]{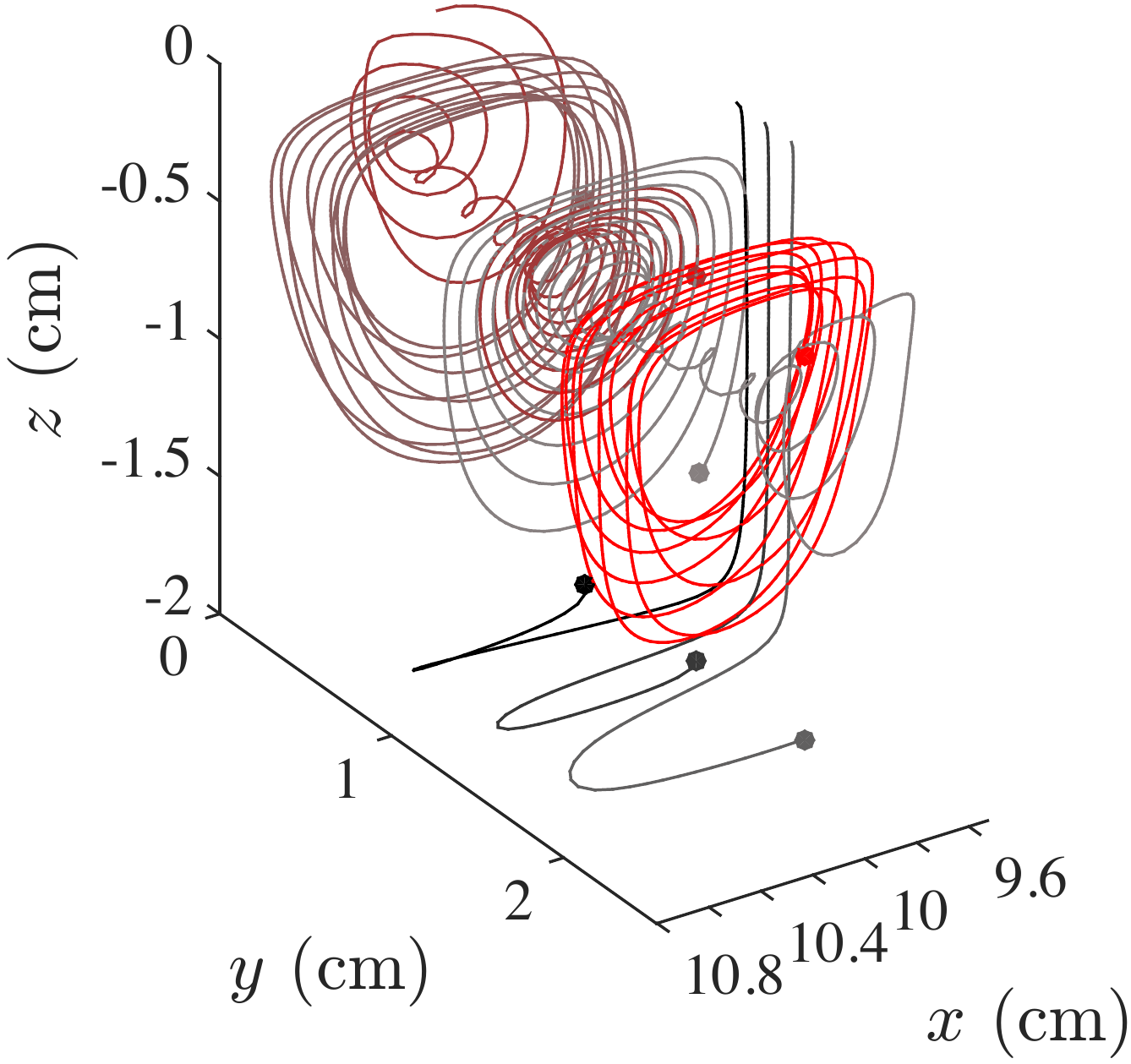}
\par\end{center}%
\end{minipage}
\par\end{centering}

\protect\caption{Trajectories of particles starting from various points of the plane
$x=9.80$ cm (central cell) for $m=7$, $f/2=5.4$ Hz and $A=0.4\,\mathrm{cm^{2}/s}$
with fully contaminated interface. Left: viewed from the face $x=0$.
Right: slanted view. Each color denotes the trajectory of the same
particle in left and right plots. The starting points of the paths
are represented by disks.\label{fig:trajectoires-C-centre}}
\end{figure}

Finally, we depict the motion of fluid particles inside the bulk in
the low amplitude limit, by plotting streamlines outside the $y$
midplane. We restrict ourselves to the lowest amplitude of patterns
$A=0.4\,\mathrm{cm^{2}/s}$ where the approximations necessary for
the theory of streaming are largely satisfied. Figures \ref{fig:trajectoires-NC-extremites}--\ref{fig:trajectoires-NC-centre}
represent the trajectories of particles in the uncontaminated case
for two longitudinal planes, $x=0.28$ cm and $x=9.8$ cm (respectively
close to a side wall and close to the $x$ mid plane). We show the
same data for the fully contaminated interface in figures \ref{fig:trajectoires-C-extremites}--\ref{fig:trajectoires-C-centre}.

In all the figures \ref{fig:trajectoires-NC-extremites}--\ref{fig:trajectoires-C-centre},
the $y$ midplane, which is a symmetry plane of the problem, plays
the role of separatrix between both halves of the domain: particles
starting within a half stay indefinitely there. The $x$ midplane
plays the same role of separatrix. A more detailed inspection shows
that all the planes of equation $x=m'L_{x}/(2m)$ are separatrices
too, $m'$ being an integer, except those at $m'=1$ and $m'=m-1$
in the uncontaminated case (compare the span of the trajectories in
the $x$ direction in figure \ref{fig:trajectoires-NC-extremites}
and \ref{fig:trajectoires-NC-centre}--\ref{fig:trajectoires-C-centre}).
Hence, the motion of a fluid particle remains confined to small portions
of the box of length $L_{x}/(2m)$ and width $L_{y}/2$, depending
on its starting point. The separatrices are also visible in figures
\ref{fig:LC-NC}--\ref{fig:LC-C} at the locations of vertical streamlines
as well as in figure \ref{fig:v_z=00003D-1} separated by the orthogonal
planes of null $u$ or $v$ components. It is important to emphasize
that odd $m'$ values correspond to nodes of the interface vibration
whereas even $m'$ values, to the antinodes. 

Outside the separatrices, the complexity of the trajectories is mainly
illustrated by the paths on the upper part of the domain. The loops
formed by the trajectories differ in figure \ref{fig:trajectoires-NC-centre},
displaying orthogonal directions to those of figures \ref{fig:trajectoires-NC-extremites},
\ref{fig:trajectoires-C-extremites} and \ref{fig:trajectoires-C-centre}.
In this case, the motion displays a transverse components tronger
than the longitudinal one. This is due to the absence of constraints
on the velocity field at the top (Neumann conditions) and the distance
to the lateral boundaries. The separatrices disappear when the amplitude
increases and the patterns become disordered. The trajectories of
all fluid particles span to the whole domain.

\section{Summary and discussion\label{sec:Discussion-2}}

In this work, we have studied the steady streaming flows sustained
by longitudinal Faraday waves in a rectangular container. We have
faced this problem by performing experiments, developing a theoretical
framework and running numerical simulations. 

In the experiments conducted in \S\ref{sec:Experiments} we report
the observation of streaming flows with a well-resolved spatial structure.
These patterns appear at any wavenumber, but their morphology may
vary among three types. Type I corresponds to well-defined counter-rotative
rolls, Type II, to moustache-like patterns and Type III, to irregular
patterns. All these patterns have been observed when performing stroboscopic
measurements of tracers, which naturally filters off the periodic
motion of standing surface waves. We have obtained the flow streamlines
by superposing the sequence of images, and the flow velocity field
by performing PIV analysis. Results were summarized in a bifurcation
diagram for the streaming vertical velocity as a function of the wave
amplitude (figure \ref{fig:Bifurcation-diagram}). All the data collapses
into two branches distinguishing type-I patterns from type-II and
type-III ones. Furthermore, the type-I and II branches coexist in
a common range of forcing parameters, which means that identical Faraday
waves can sustain qualitatively different streaming flows. Other parameters,
such as the wavelength or the bifurcation criticality do not show
any important qualitative role. 

We developed a theory for the streaming due to oscillatory flows in
\S\ref{sec:Theory}, extending the results of Batchelor to three
dimensional configurations. The theory is based on boundary layers
generated by tangential oscillating flows in the vicinity of rigid
walls. We show that boundary layers slowly induce a vorticity into
the bulk until a steady state \textemdash the streaming flow\textemdash{}
is reached. Starting from Navier-Stokes equations, we split the fluid
domain in two regions: the bulk, where viscosity is neglected, and
the boundary layer where viscosity plays a major role. Then, using
asymptotic analysis, we find the right hierarchy of equations which
allows us to match the bulk and the boundary layer flows at the junction.
At this point we introduced a flow solution having both an oscillatory
and a steady component. By performing time averages along a period
of oscillation, we extract the steady component of the flow in the
boundary layer. Equations show that the streaming inside the boundary
layer induces a net streaming in the junction with the bulk, which
can then be used as a matched boundary condition for the bulk region.
In the two-dimensional case, these conditions are consistent with
those obtained by \citet{Batchelor2000an_introduction}. Finally,
the theory explains that the boundary layers represent the main source
of streaming in the vibrating fluid layer, (corrections due to nonlinearities
in the bulk are of higher order). Focused on our experiments, we have
applied this general theory to find the streaming induced by the rigid-walls
in longitudinal Faraday waves in a rectangular container. On the other
hand, complementary boundary conditions at the free surface have been
fixed accounting for clean and fully contaminated cases.

Using the theoretical results, we performed direct numerical simulations
aiming to the steady component sustained by Faraday waves. This has
been applied to a rectangular domain with a single fluid, where the
oscillating Faraday wave main flow only couple through the matched
boundary conditions previously deduced. From the simulations we have
observed various types of streaming patterns, depending on the surface
conditions. In the case of fully contaminated surface, we have obtained
counter-rotative structures that strikingly resemble type-I patterns
observed in our experiments. For a clean surface, similarities with
experiments have also been found although in a less conclusive way.
Considering conditions of intermediate contamination \citep{2005JFM...546..203M}
may improve the agreement with patterns of type-II. Irregular patterns
(type III) have been observed in simulations when the forcing amplitude
is increased for any surface condition. Then we have computed a bifurcation
diagram analogous to the experimental one. We have observed that quantitatively,
results obtained in simulations are compatible with experimental measurements.
Also, our results with a contaminated surface and clean surface present
distinguishable vertical velocities. However, the collapse of the
curves found in experiments is not observed in the numerical results,
an issue which deserves further investigation. Then we have focussed
on whether the transversal effects are truly negligible. The comparison
between three-dimensional streaming flows and two-dimensional ones
shows that the walls in the third dimension have an important effect
on the streaming flow, remarkably large compared to their effect on
surface waves. By computing the horizontal components of the velocity
fields and the trajectories of several fluid particles, we have concluded
that a two-dimensional approximation is not enough to catch the richness
of the flow beneath Faraday waves.

The original investigation of \citet{1831RSPT..121..299F} and recent
experimental attempts \citep{Falkovich:2005ei,Sanl:2014hs} have naturally
risen the question about how streaming flows may influence the motion
of floaters or heavy particles. Despite it is out of the scope of
the present investigation, our results show that streaming flows are
another mechanism leading to slow time scale motion of such particles.
Contrary to weight and bouyancy that can only induce vertical motion,
the average drag induced by the streaming does contribute with horizontal
net forces exerted on the particles, even when they are constrained
to the bottom or top of the fluid. In particular, the discussed set
of separatrices act as basins of repulsion and attraction for particles,
whose locations depend on both the wavelength of the Faraday waves
and the contamination of the surface. This interesting perspective
will be the subject of a rich field for further exploration.

Finally, the comparison between numerical and experimental results
presented in this work poses an interesting issue. While the numerical
results point to the free surface contamination as the only variable
dramatically affecting the streaming flow, the experimental results
show that identical Faraday waves (same fluid under same conditions)
may mask qualitatively different streaming flows. The link will strikingly
require that the same liquid can display different contamination degrees,
which is contrastable with the general assumption in the literature
that the degree of contamination is a static feature of the flow.
This provides a hint that contaminants are not just passively dragged
by the streaming flow but also exert a feedback on it: a feature which
enriches the streaming problem even further.

\section*{Acknowledgements}

We are grateful to Héctor Alarcón for fruitful discussions. LG was
supported by Conicyt/Becas Chile de Postdoctorado 74150032. PG was
supported by Conicyt/Fondecyt Postdoctorado 3140550. NP was supported
by Conicyt/Fondecyt Postdoctorado 3140522. HU was supported by Fondecyt
11130450. The laser, the fast camera and some other parts of the setup
were purchased under the funding program Conicyt AIC-43. Powered@NLHPC:
This research was partially supported by the supercomputing infrastructure
of the NLHPC (ECM-02).


\end{document}